\documentclass[utf8]{frontiersFPHY} % for Physics and Applied Mathematics and Statistics articles

\setcitestyle{square} % for Physics and Applied Mathematics and Statistics articles
\usepackage{url,hyperref,lineno,microtype,subcaption}
\usepackage[onehalfspacing]{setspace}

\def\keyFont{\fontsize{8}{11}\helveticabold }
\def\firstAuthorLast{Davide Lazzati} %use et al only if is more than 1 author
\def\Authors{Davide Lazzati\,$^{1,*}$}
\def\Address{$^{1}$Department of Physics, Oregon State University, Corvallis , OR, USA }
\def\corrAuthor{Corresponding Author}

\def\corrEmail{davide.lazzati@oregonstate.edu}

\begin{document}
\onecolumn
\firstpage{1}

\title[SGRBs after GW170817]{Short duration gamma-ray bursts and their outflows in light of GW170817} 

\author[\firstAuthorLast ]{\Authors} %This field will be automatically populated
\address{} %This field will be automatically populated
\correspondance{} %This field will be automatically populated

\extraAuth{}% If there are more than 1 corresponding author, comment this line and uncomment the next one.
%\extraAuth{corresponding Author2 \\ Laboratory X2, Institute X2, Department X2, Organization X2, Street X2, City X2 , State XX2 (only USA, Canada and Australia), Zip Code2, X2 Country X2, email2@uni2.edu}

\maketitle

\begin{abstract}

%%% Leave the Abstract empty if your article does not require one, please see the Summary Table for full details.
\section{}
The detection of GW170817, it's extensive multi-wavelength follow-up campaign, and the large amount of theoretical development and interpretation that followed, have resulted in a significant step forward in the understanding of the binary neutron star merger phenomenon as a whole. One of its aspects is seeing the merger as a progenitor of short gamma-ray bursts (SGRB), which will be the subject of this review. On the one hand, GW170817 observations have confirmed some theoretical expectations, exemplified by the confirmation that binary neutron star mergers are the progenitors of SGRBs. In addition, the multimessenger nature of GW170817 has allowed for gathering of unprecedented data, such as the trigger time of the merger, the delay with which the gamma-ray photons were detected, and the brightening afterglow of an off-axis event. All together, the incomparable richness of the data from GW170817 has allowed us to paint a fairly detailed picture of at least one SGRB. I will detail what we learned, what new questions have arisen, and the perspectives for answering them when a sample of GW170817-comparable events have been studied.      

\tiny
 \keyFont{ \section{Keywords:} gamma-ray bursts, relativistic astrophysics, hydrodynamics, transient sources, gravitational waves, binary mergers}
 
\end{abstract}

\section{Introduction}

Gamma-ray bursts (GRBs) are some of the most energetic explosions in the present day Universe, characterized by the release of large amounts of energy, within a few milliseconds to tens of seconds, resulting in the acceleration of relativistic outflows and the release of high-energy photons \cite{Fishman1995,Piran1999,Meszaros2002,Gehrels2009,Kumar2015}. They can be divided in at least two classes, based on the duration of their prompt phase, in which their emission is concentrated in the hard X-ray and gamma-ray bands and is characterized by fast variability \cite{Kouveliotou1993}. Long duration GRBs last two seconds or more, while short duration GRBs (SGRBs) last between a few milliseconds and two seconds. Alternative classifications have also been introduced, considering, e.g., short GRBs with extended emission \cite{Norris2006,Norris2010,Dainotti2010,Barkov2011,Dainotti2017}, or attempting a more physical classification based on inferred progenitor properties \cite{Lu2010,Bromberg2013}.

In the last two and a half decades, the study of GRBs has concentrated on long duration GRBs, and a general consensus has grown around a model in which these events are associated with the collapse of the core of massive stars \cite{Woosley2006}. While the collapse of most massive stars would ignite a core-collapse supernova, those that are fastly spinning and metal poor could also trigger a long duration GRB, powered by a compact central engine \cite{Woosley1993,Woosley2006b,Yoon2006}.  Whether the central engine is a fastly spinning, highly magnetized neutron star (NS) \cite{Bucciantini2008,Bucciantini2009,Metzger2011} or an accreting black hole (BH) \cite{Woosley1993,Lee2000,Lei2013} is the matter of open debate.

The interest on SGRBs had increased in the last decade, initially as a consequence of the launch of the Fermi satellite, which had a higher efficiency for detecting and localizing them compared to its predecessors \cite{Meegan2009}. More recently, the theoretical expectation that SGRBs had to be associated with the merger of binary NS systems (or, perhaps, system made by a BH and a NS) \cite{Belczynski2006,Lee2007,Fong2013a,Giacomazzo2013,Ruiz2016,Ciolfi2018} has made them the expected and highly anticipated high-frequency counterparts of gravitational wave sources \cite{Nakar2006,Kiuchi2010,Schutz2011}. Such expectations were supported by energetic and temporal arguments. Powering a GRB requires a large amount of energy, comparable to the rest mass of a stellar object converted to energy. In addition, said energy needs to be released in a matter of a fraction of a second, at least for SGRBs. Naked compact objects (NS and BH) are the only available candidates that can offer the required energy within a region of less than a light second. However, isolated NS and BH are unlikely progenitors, since some catastrophic event needs to take place to cause the sudden release of a large fraction of their total energy. Binary mergers are therefore a natural candidate, when at least one of the two members is a NS, since a binary BH system would merge in a bigger BH that would swallow all the matter and energy, instead of ejecting them as a relativistic outflow\footnote{Some have suggested, however, that even binary BH mergers could produce a weak electromagnetic transient, under certain conditions \cite{Perna2016,Liu2016,Zhang2016}.}

All these expectations were confirmed by the detection of GW170817 \cite{Abbott2017a} and its associated gamma-ray burst GRB170817A \cite{Abbott2017c,Goldstein2017,Savchenko2017,Zhang2018}, afterglow, and kilonova \cite{Abbott2017b,Coulter2017,Cowperthwaite2017,Pian2017,Kasen2017,Troja2017,Smartt2017,Tanvir2017,Hallinan2017,Soares-Santos2017,Evans2017,Arcavi2017,Margutti2017,Mooley2018a,Margutti2018,Alexander2017,Mooley2018b,Valenti2017,Haggard2017,Lyman2018,Ghirlanda2019,Davanzo2018,Resmi2018,Lamb2019,Kim2017,Abdalla2017,Siebert2017,Villar2018}. In this contribution I will review the key observations of GW170817 as a SGRB (also known as GRB170817A), the questions that were answered, and the new ones that were spurred, and briefly discuss what more insight is expected from the detection of more systems akin to GW170817 in future GW observing runs.

\section{Before the prompt emission}
\label{sec:before}

In this section I will review the physics of the SGRB outflow before the prompt emission phase begins, as it happened in GW170817. First of all, there is little doubt that the GW signal of GW170817 came from a binary compact merger, and that the masses of the two compact objects are compatible with being NS \cite{Abbott2017a,Abbott2017b,Abbott2017c}. The GW signal by itself does not allow to distinguish between NSs and BHs, but the richness of the electromagnetic signal that followed requires the presence of baryonic matter, and therefore at least one of the two components of the binary had to be a neutron star. Most likely they were both NSs \cite{Coughlin2019}.

\bigskip
\subsection{The time delay}

\begin{figure}[h!]
\begin{center}
\includegraphics[width=\textwidth]{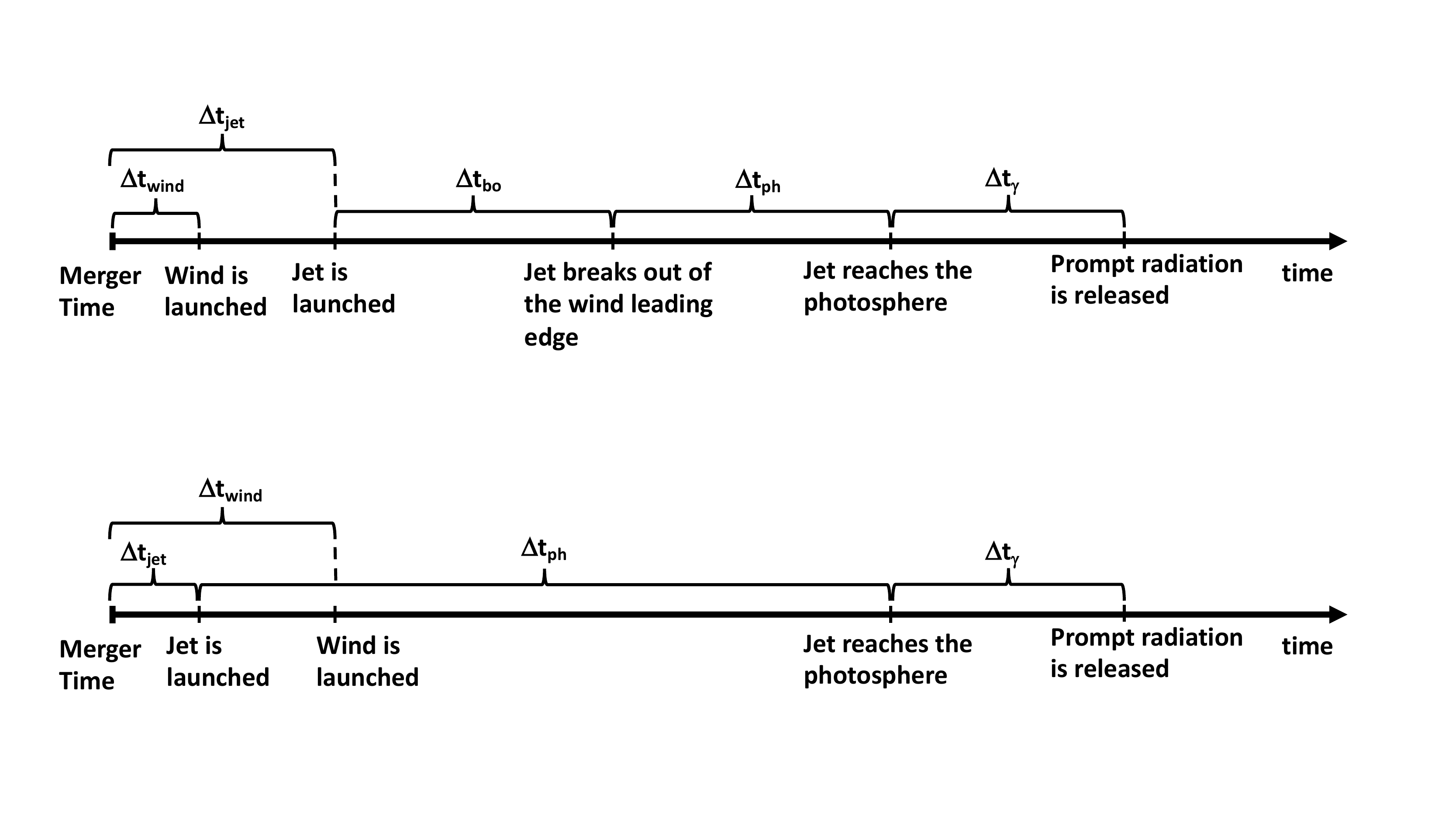}% This is a *.eps file
\end{center}
\caption{The two possible timelines with all the phases that may contribute to the detected delay $\Delta{t}_{\rm{GW-\gamma}}$. Due to the presence of a structured outflow, GW170817 most likely followed the top timeline. The relative contribution of the various phases is a matter of debate, but consensus is growing around  $\Delta{t}_{\rm{wind}}<\Delta{t}_{\rm{jet}}\ll1$~s, $\Delta{t}_{\rm{bo}}\ll1$~s, $\Delta{t}_{\gamma}\sim0$, and  $\Delta{t}_{\rm{ph}}\sim\Delta{t}_{\rm{GW-\Gamma}}$ .} \label{fig:timeline}
\end{figure}

Besides the identification of the progenitor, a very important piece of information that GW170817 provided is the merger time, which allowed for the measuring of the time delay between the GWs and the gamma-ray signals. This delay, which we indicate as $\Delta{t}_{\rm{GW-\gamma}}$ can be due to several reasons, as detailed below and shown in Figure~\ref{fig:timeline} \cite{Granot2017,Lin2018,Zhang2019a,Lazzati2020,Lucca2020}.

\begin{itemize}
    \item {\bf Engine Delay|} While the time of the merger is the earliest time at which the jet from the central engine can be produced, there is the possibility of some delay \cite{Cook1994,Lasota1996,Vietri1998,Ciolfi2015}. Such delay is difficult to predict theoretically
    but can be likely due either to the need of a transition in the engine itself or to the need of amplifying the magnetic field to a value large enough to launch a jet. The former can be quite long, up to years, and usually invokes a metastable, fastly spinning NS that collapses into a black hole when its rotation period is increased by either internal or external torques. We indicate this delay time as $\Delta{}t_{\text{eng}}$.
    \item {\bf Wind Delay|} Owing to the detection of a kilonova (KN) and an off-axis SGRB from a structured outflow, we know that GW170817 ejected a non-relativistic wind. There can be a delay in launching such a wind as well, and we indicate it as $\Delta{}t_{\text{wind}}$. It should be noted, however, that this delay can in principle be negative since the neutron star surfaces are tidally shredded in the last few orbits before the merger.
    \item {\bf Breakout delay|} If the wind is ejected before the jet, then the jet has to propagate through the wind. The propagation happens at sub-relativistic speed, causing a delay of the head of the jet with respect to the GW signal that travels at the speed of light \cite{Matzner2003,Morsony2007,Bromberg2011,Lazzati2019}. We indicate the time it takes for the jet head to cross the wind as $\Delta{t}_{\rm{bo}}$.  The jet-wind interaction also causes the development of  a cocoon \cite{Ramirez-Ruiz2002}, confined by the surrounding wind. This leads to the development of a structured outflow that maintains a bright core but develops wide, energetic wings at large polar angles \cite{Lazzati2017a,Lazzati2017b}.
    \item {\bf Photospheric delay|} After the outflow has broken out of the leading edge of the wind, it needs to propagate out to the photospheric radius. At this point the jet becomes transparent and the necessary conditions for the release of the prompt gamma-ray radiation are met. We indicate the delay due to the propagation from the break out radius to the photospheric radius as $\Delta{t}_{\rm{ph}}$.
    \item {\bf Dissipation delay|} While at the photospheric radius the prompt emission can be radiated, it does not mean it is. In some models, such as the popular internal shock synchrotron model, the outflow needs to propagate out to the internal shock radius before the bulk energy of the flow is dissipated and turned into radiation. We indicate this additional delay as $\Delta{t}_{\gamma}$.
\end{itemize}

For the first time, a measurement of the sum of all these possible delays was available for GW170817 \cite{Abbott2017c}. The prompt gamma-ray radiation was detected with a delay 
$\Delta{t}_{\rm{GW-\gamma}}\simeq1.75$~s. Several attempts have been made to constrain the various individual contributions, but a general consensus has not been achieved \cite{Gill2019,Shoemaker2018,Hamidani2020,Beniamini2020}. A few robust inferences can however be made \cite{Lazzati2020}. Overall, the measured delay was fairly small, since GW170817 ejected a significant amount of energy towards the observer but its Lorentz factor could be at most moderate ($\Gamma<7$) \cite{Beniamini2020aa}. These combine to a large photospheric radius and a photospheric delay 
\begin{equation}
    \Delta{t}_{\rm{ph}}\sim \frac{R_{\rm{ph}}}{c\Gamma^2}=1.4 
    \frac{R_{\rm{ph}}}{2\times 10^{12}~\rm{cm}} 
    \left(\frac{7}{\Gamma}\right)^2 \;\; \rm{s}
\end{equation}
The photospheric delay therefore had to contribute to a sizable part of the delay, and it represent a lower limit to the observed delay, since any other non-photospheric emission mechanism would require a longer delay (this allows to use the above equation to put a Lower limit on $\Gamma$ \cite{Beniamini2020}). The wind delay, if there was any, had to be smaller than the jet delay, so that the jet-wind interaction could generate a structured outflow, as requested for modeling the afterglow emission. For the same reason, the jet delay itself could be fairly small but could not be null. Finally, the breakout and dissipation delays had to be small in order to accommodate the large expected photospheric delay. Note, however, that the prompt emission spectrum had a non-thermal shape, a property that is not expected form a simple photospheric emission model (see Section~\ref{sec:prompt} for a more thorough discussion). 

\bigskip
\subsection{The shaping of the outflow}

GW170817 was also the first GRB for which evidence of a structured outflow could be unequivocally determined.
The structure of the outflow could be intrinsic, as the jet itself could have been launched with a non-uniform polar structure \cite{Aloy2005,Kathirgamaraju2019}. However, the relatively large energetics of GW170817 in gamma-rays and the shape of its afterglow lightcurve (see Section~\ref{sec:aglow}) suggest a wide structure, most likely brought about by the jet interaction with the wind from the merger \cite{Lazzati2019,Salafia2020}. 

A typical SGRB jet with isotropic equivalent energy $E_{\rm{iso}}=10^{53}$~erg and asymptotic Lorentz factor $\eta=100$
has a baryon rest mass $M_0=E_{\rm{iso}}/\eta{}c^2\sim10^{-4}M_\odot$. If it encounters a wind mass $M_{\rm{wind}}\ge{}M_0/\eta\sim10^{-6}M_\odot$ it is shocked and the velocity of propagation of its head is slowed until the working surface of the jet head is in causal contact, allowing for the wind material and the shocked jet material to move to the side instead of accumulating in front of the jet and thereby slowing it down \cite{Matzner2003,Morsony2007,Bromberg2011,Lazzati2019}. As a consequence, a high-pressure cocoon inflates around the jet, composed by partially mixed jet and wind material. As the jet breaks out of the wind leading edge, the cocoon loses the confining effect of the wind material and is released. Since it has large pressure, it accelerates creating a broad structure around the jet with decreasing energy and Lorentz factor for increasing polar angle. This process therefore turns a collimated jet into a structured outflow. It requires a small wind mass that is well below the expected amount of baryons ejected in a binary NS merger. The cocoon structure can be studied analytically, by enforcing pressure balance between the jet, cocoon, and wind material at their respective contact surfaces, or through numerical simulations. Despite its importance for predicting burst/merger observability and understanding the structure and composition of the merger wind and jet, the polar profile of the outflow is highly debated. Analytic functions ranging from Gaussian, power-law, and exponential have been tested, and even numerical simulations do not provide an unequivocal answer \cite{Nagakura2014,Murguia-Berthier2014,Lazzati2017b,Murguia-Berthier2017,Murguia-Berthier2017a,Xie2018,Kathirgamaraju2019,Wu2018,Duffell2018,Granot2018a,Geng2019,Gill2019c,Hamidani2020,Hamidani2020a,Murguia2020,Takahashi2020,Takahashi2020a}.

\section{The prompt emission}
\label{sec:prompt}

Approximately 1.75 seconds after the GW chirp, a gamma--ray pulse was observed by both the Fermi and INTEGRAL satellites from a position compatible with the direction from which the GWs arrived \cite{Goldstein2017,Savchenko2017}. The pulse was made by an initial spike of about half a second followed by a broader, less intense tail, for an overall duration of $\sim2$~s. Two characteristics make this gamma-ray pulse different from the population of previously observed SGRBs: it is markedly less energetic than an average cosmological SGRB and, given its energetics, it has a very high peak frequency \cite{Fong2015}. As a matter of fact, the detection itself was surprising because the chance of having a SGRB jet pointing along the line of sight for the first GW-selected binary merger was expected to be small \cite{Metzger2012,Ghirlanda2016}. That is because the amplitude of the GWs depend only mildly on the orientation of the binary, while the intensity of the radiation from a narrow, relativistic jet drops quickly for any line of sight outside the jet itself. Such an expectation was based, however, on the properties of a narrow jet and not on the possibility that the jet-wind interaction would cause a structured outflow to form. Predictions from models with structured outflows had indeed shown that, for moderately large off-axis angles, a detectable signal would be expected from a GW-detected merger \cite{Lazzati2017a,Lazzati2017b}. A similar effect might be responsible for X-ray flashes, when a long duration GRB is seen off-axis \cite{Yamazaki2002,Yamazaki2003}.

\begin{figure}[h!]
\begin{center}
\includegraphics[width=\textwidth]{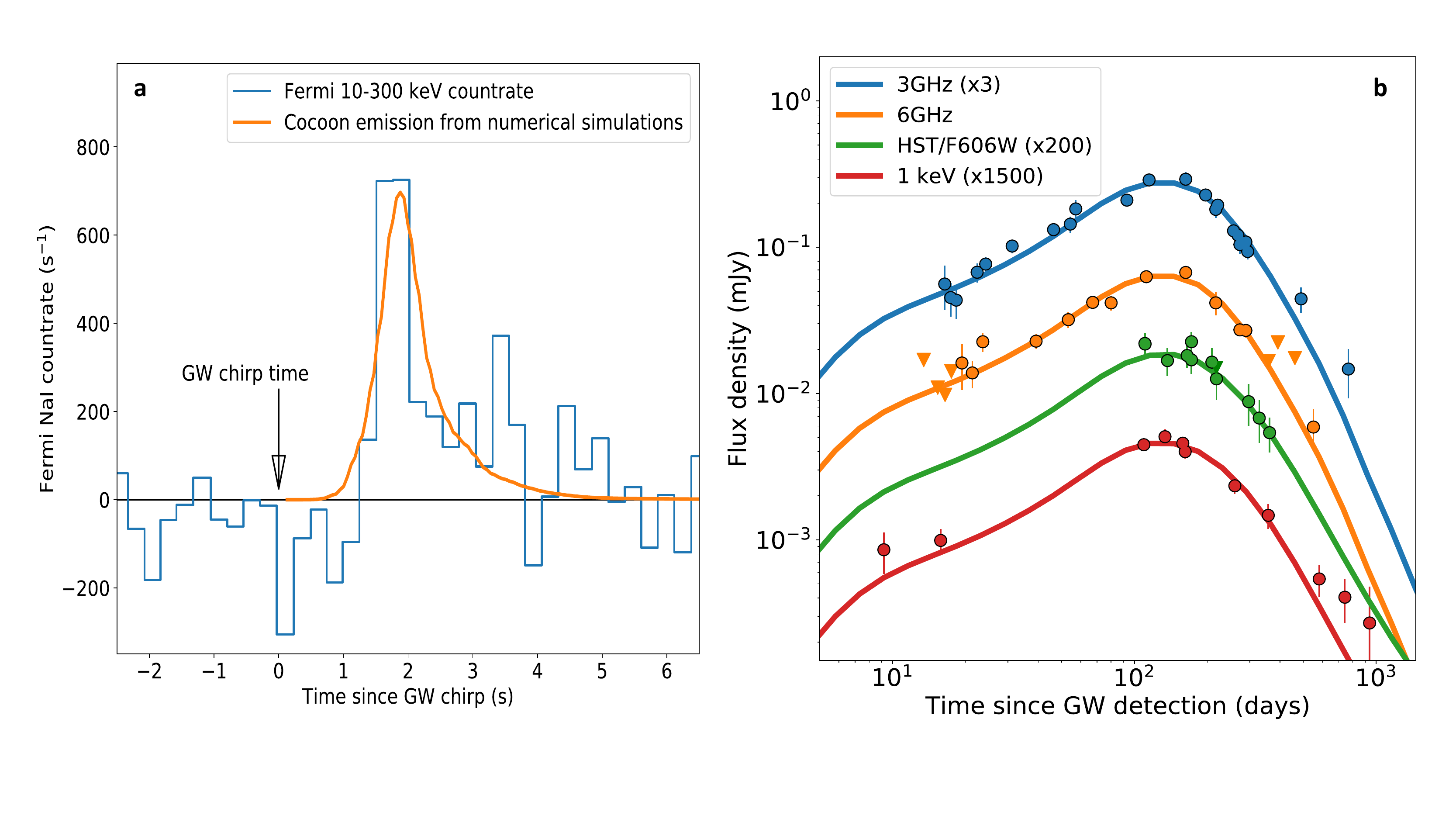}% This is a *.eps file
\end{center}
\caption{Left panel (panel a): the prompt emission of GW170817. The blue step-line shows the Fermi data \cite{Goldstein2017}, while the orange solid line 
is the prediction from a theoretical simulation that assumes a structured outflow from the jet-wind interaction \cite{Lazzati2017b}. The radiation is assumed to be released at the photosphere. Right panel (panel b): Afterglow of GW170817. Symbols with error-bars show observations in the radio, optical, and X-ray bands. Solid lines show the best fit result for an afterglow model with a structured outflow and an observer located at $\theta_o=35^\circ$ from the line of sight. Additional data at different radio frequencies were used to constrain the model, but only two radio bands are shown for clarity. Adapted from \cite{Makhathini2020}} \label{fig:composite}
\end{figure}

The structured outflow model was successful at predicting that a SGRB would be detectable even at large off-axis angles \cite{Lazzati2017a,Lazzati2017b}. It correctly predicted the off-axis burst energetics and its duration. It could also successfully explain the detected delay between the GWs and the $\gamma$-rays. A comparison between the Fermi data and the bolometric photospheric emission \cite{Lazzati2017b} is shown in the left panel of Figure~\ref{fig:composite}. The one aspect of GW170817 that cannot be accounted for by the simple photospheric cocoon emission is the $\gamma$-ray spectrum of the prompt emission. At least in first approximation, the photosphere of an off-axis structured outflow is expected to produce a thermal pulse with temperature \cite{Lazzati2017a,Lazzati2017b}
\begin{equation}
    T_{\rm{obs}}\simeq \left(\frac{L\Gamma^2}{4\pi\sigma R_{\rm{ph}}^2}\right)^\frac{1}{4}=
    10^7 \left(\frac{L}{10^{47} \rm{erg}}\right)^\frac{1}{4}
    \left(\frac{\Gamma}{100}\right)^\frac{1}{2}
    \left(\frac{10^{12} \rm{cm}}{R_{\rm{ph}}}\right)^\frac12 \quad \rm{K}
\end{equation}
which would produce a spectrum peaked at a few KeV, in severe tension with the observed peak frequency at $\sim150$~keV \cite{Goldstein2017}. This is due to the fact that the cocoon, which energized the outflow at large off axis, is not expected to be radially structured, and therefore no significant dissipation is expected to occur around the photospheric radius, differently from the photospheres of long GRBs \cite{Lazzati2009,Parsotan2018}. One possible explanation is that the prompt radiation was due to an external shock \cite{Veres2018}. However, given the low Lorentz factor and low interstellar medium densities expected in the surroundings of GW170817, the timing of the prompt emission, less than two seconds after the launching of the jet, is difficult to explain. Alternatively, the prompt emission could be due to the breakout of the cocoon from the leading edge of the wind \cite{Kasliwal2017,Gottlieb2018,Bromberg2018,Nakar2018a}. The shock breakout model can explain the energetics and spectrum of the prompt emission \cite{Nakar2012x} but requires a finely tuned setup in which the wind is very fast, so that it can reach a large enough radius at the breakout time. The origin of the prompt emission spectrum is therefore not  been explained in a completely satisfactory way, yet \cite{Pozanenko2018,Matsumoto2019,Ioka2019,Meng2018,Kisaka2018}. The observation of more SGRBs from GW-detected mergers will  offer further observational constraints to shed light on this remaining riddle.

\section{The afterglow}
\label{sec:aglow}

The afterglow of GW170817 had its own share of unique features. To begin with, it was not detected for more then a week, until it was bright enough to be seen first in X-rays \cite{Troja2017,Margutti2017} and, at around the two weeks mark, in radio waves \cite{Troja2017,Hallinan2017}. The detection of the afterglow at optical wavelengths had to wait for the dimming of the associated kilonova, and was performed only around day 110 with the Hubble Space Telescope \cite{Lyman2018}. Such late appearance of an afterglow is unprecedented, since the typical behavior is that the afterglow peaks very early, minutes to hours after the burst, and only dims with time afterwards \cite{vanParadijs2000,Nousek2006}. 
A second unique feature of the afterglow of GW170817 was that, even after it was detected, it sustained a slow brightening at all wavelengths \cite{Margutti2017,Mooley2018a,Ruan2018,Troja2019}, eventually peaking $\sim150$ days after the GW detection and dropping in luminosity steeply afterwords \cite{Dobie2018,Makhathini2020} (see the right panel of Figure~\ref{fig:composite}). 

The outflow from GW170817 along the direction towards Earth was under-energetic by a factor 10000 to 100000 times with respect to a typical SGRB \cite{Fong2017}. An outstanding question was therefore whether GW170817 had a misaligned, SGRB-like jet pointing in a different direction or not \cite{Lazzati2018,Salafia2018,Mooley2018b}. If it did, then the identification of the SGRB progenitors with binary NS mergers would be secured. If if did not, then what GW10817 was associated with would be a new class of dim, possibly isotropic, $\gamma$-ray transients. Unfortunately, telling whether a misaligned relativistic jet is present is not easy, since all the radiation is relativistically beamed away from the line of sight. The slow but steady brightening was shown to be consistent with the presence of a jet, its energy contribution along the line of sight growing with the deceleration of the external shock \cite{Lazzati2018,Granot2018,Nakar2018a,Lamb2018a,Finstad2018,Xiao2017,Oganesyan2020,Fraija2019,Fraija2019b,Beniamini2020aa,Decolle2018}. However, a radially stratified spherical outflow could reproduce the observations as well, albeit at the price of adding a never  observed before component to the models \cite{Mooley2018a,Nakar2018,Nakar2018a,Li2018,Salafia2018}. Some evidence in favor of a jet was provided by the steep post-peak decay at all wavelengths \cite{Alexander2018,Mooley2018c,Nynka2018,Hajela2019,Lamb2018c,Fong2019,Jin2018}. In addition, it was soon realized that either a relatively large linear polarization \cite{Gill2018} or a small but detectable proper motion of the radio transient could potentially give the final clue. Both observations were carried out. Polarization turned out to be small \cite{Corsi2018}, and only an upper limit of $12$ per cent was obtained, still consistent with either explanation. Long baseline radio interferometry turned out to be the key. In one experiment, a small but significant proper motion was detected \cite{Mooley2018b}, while in a second experiment the radio source was confirmed to be point-like \cite{Ghirlanda2019}. Both these characteristics are incompatible with a spherical expansion. In the future, the detection of the counter-jet emission might give additional evidence \cite{Yamazaki2018}.

To date, despite the very high quality of the available data, the unique afterglow of GW170817 can be modeled successfully with the good old external shock synchrotron model \cite{Meszaros1997,Sari1998}, with the only required addition of considering off-axis observers \cite{Granot2002} and allowing for some structure in the polar direction \cite{Lazzati2017b,Lazzati2018}. The type of polar stratification is not univocally constrained, since Gaussian, power-law, and exponential profile seem all to give an adequate fit to the data \cite{Troja2018,Lazzati2018,Xie2018,Ghirlanda2019}. Numerical simulations are also ambiguous, different codes yielding different polar structures, including the three mentioned above \cite{Nagakura2014,Murguia-Berthier2014,Lazzati2017b,Murguia-Berthier2017,Murguia-Berthier2017a,Xie2018,Kathirgamaraju2019,Wu2018,Duffell2018,Granot2018a,Geng2019,Gill2019c,Hamidani2020,Hamidani2020a,Takahashi2020,Takahashi2020a}. Constraints can be obtained from the lack of a large populations of cosmological off-axis bursts \cite{Beniamini2019a}. More observations and further theoretical work are needed to pin down this important aspect that has implications not only on the detectability of bursts but also on the nature of the inner engine and the composition of the ejected jet and wind.

\section{Summary, discussion, and a look at the future}

GW170817 was a rich event, a cornerstone detection in our understanding of SGRBs. It confirmed that binary NS mergers are the progenitor of at least some short bursts, it showed us that the top-hat jet model is woefully inadequate for describing the relativistic outflows of SGRBs (and possibly long duration GRBs as well) and it gave us, for the first time ever, a measure of the trigger time and of the delay between the launching of the jet and the detection of the prompt emission radiation.

We now know that the burst associated with GW170817 was a fairly canonical SGRB \cite{Salafia2019}, with a powerful relativistic jet that, after interacting with the merger wind, turned into a structured outflow \cite{Lazzati2017b,Lazzati2018}. Our line of sight lied somewhere between 15 and 35 degrees away from the jet axis, the lower value obtained by high resolution radio imaging \cite{Mooley2018a,Ghirlanda2019}, while the larger value being favored by multi-band afterglow modeling and ejecta considerations \cite{Mandel2018,Zou2018,Lazzati2018}. The prompt emission was powered by an energetic cocoon inflated by the interaction of the jet with the merger wind. The gamma-ray radiation was likely released at or near the photosphere, either by a shock breakout \cite{Nakar2012x,Kasliwal2017,Gottlieb2018} or by other non-thermal mechanisms \cite{Savchenko2017,Veres2018}. The external shock developed later than usual due to the lower than customary Lorentz factor of the outflow along the line of sight and the afterglow was unusual, characterized by an initial increase in luminosity that lasted for a few months before peaking and beginning a steep declining phase. This behavior is understood to be due to the structure of the outflow, characterized by a polar stratification with a steep decline as a function of angle in both the energy per unit solid angle and the Lorentz factor.

Despite the large amount of observational evidence that allowed us to paint a detailed picture of the dynamic of the relativistic ejecta of GW170817 and their electromagnetic signatures, some questions remain open. First, we do not know the nature of the compact object that launched the relativistic jet. It could have been either a meta-stable NS or a BH, and consensus in this respect hasn't been reached \cite{Metzger2018,Piro2019,Pooley2018,Abedi2019}. A related mystery is the origin of the observed 1.75~s delay between the GW and the prompt emission. As discussed in Section~2.1 the delay is the sum of many components and it is unclear which dominates, or if several of them have comparable magnitude. Since the photospheric delay is strongly dependent on the viewing angle, observation of several SGRBs from a diverse set of angles will help better understand the origin of the delay. Still unclear is also the physics of the dissipation that powered the prompt emission and the prompt emission mechanism itself. Shock breakouts, internal dissipation such as internal shocks, and even external shocks have been proposed (see Section~3).

Finally, we still do not know how typical GW170817 was. The fact that most likely it originated from a binary NS merger does not exclude the possibility that some --- if not most --- SGRB are made in NS-BH mergers. It might even be that GW170817 itself was a NS-BH merger \cite{Coughlin2019,Kyutoku2020}. Re-analysis of several past bursts have yielded some support the the presence of kilonovae in their light curves  \cite{Beniamini2019,Troja2018a,Troja2019b,Lamb2019b} or similarities in their prompt emission \cite{Burns2018,vonKienlin2019}, showing that GW170817 was not unique. However, there might be cases in which the jet is not successful in breaking out of the wind leading edge, and a weaker transient would be produced \cite{Kasliwal2017,Mooley2018a,Salafia2018}. Future GW detections with the power of multimessenger observations will allow to better understand the connection between binary NS mergers, binary NS-BH mergers, and SGRBs.

\section*{Author Contributions}

DL has written this review article on his own to the best of his knowledge, he has produced all figures himself from publicly available data and codes. The references are extensive but they are by no means exhaustive.

\section*{Funding}
DL acknowledges support from NASA grants 80NSSC18K1729 (Fermi) and NNX17AK42G (ATP), Chandra grant TM9-20002X, and NSF grant AST-1907955.

%\section*{Acknowledgments}

%\bibliographystyle{frontiersinSCNS_ENG_HUMS} % for Science, Engineering and Humanities and Social Sciences articles, for Humanities and Social Sciences articles please include page numbers in the in-text citations
\bibliographystyle{frontiersinHLTH&FPHY} % for Health, Physics and Mathematics articles
\bibliography{sgrb}

\begin{thebibliography}{173}
\expandafter\ifx\csname natexlab\endcsname\relax\def\natexlab#1{#1}\fi
\expandafter\ifx\csname urlstyle\endcsname\relax
  \expandafter\ifx\csname doi\endcsname\relax
  \def\doi#1{doi:\discretionary{}{}{}#1}\fi \else
  \expandafter\ifx\csname doi\endcsname\relax
  \def\doi{doi:\discretionary{}{}{}\begingroup \urlstyle{rm}\Url}\fi \fi
\expandafter\ifx\csname selectlanguage\endcsname\relax
  \def\selectlanguage#1{}\fi

\bibitem[{{Fishman} and {Meegan}(1995)}]{Fishman1995}
{Fishman} GJ, {Meegan} CA.
\newblock {Gamma-Ray Bursts}.
\newblock {\em \araa\/} {\bf 33} (1995) 415--458.
\newblock \doi{10.1146/annurev.aa.33.090195.002215}.

\bibitem[{{Piran}(1999)}]{Piran1999}
{Piran} T.
\newblock {Gamma-ray bursts and the fireball model}.
\newblock {\em \physrep\/} {\bf 314} (1999) 575--667.
\newblock \doi{10.1016/S0370-1573(98)00127-6}.

\bibitem[{{M{\'e}sz{\'a}ros}(2002)}]{Meszaros2002}
{M{\'e}sz{\'a}ros} P.
\newblock {Theories of Gamma-Ray Bursts}.
\newblock {\em \araa\/} {\bf 40} (2002) 137--169.
\newblock \doi{10.1146/annurev.astro.40.060401.093821}.

\bibitem[{{Gehrels} et~al.(2009){Gehrels}, {Ramirez-Ruiz}, and
  {Fox}}]{Gehrels2009}
{Gehrels} N, {Ramirez-Ruiz} E, {Fox} DB.
\newblock {Gamma-Ray Bursts in the Swift Era}.
\newblock {\em \araa\/} {\bf 47} (2009) 567--617.
\newblock \doi{10.1146/annurev.astro.46.060407.145147}.

\bibitem[{{Kumar} and {Zhang}(2015)}]{Kumar2015}
{Kumar} P, {Zhang} B.
\newblock {The physics of gamma-ray bursts \&amp; relativistic jets}.
\newblock {\em \physrep\/} {\bf 561} (2015) 1--109.
\newblock \doi{10.1016/j.physrep.2014.09.008}.

\bibitem[{{Kouveliotou} et~al.(1993){Kouveliotou}, {Meegan}, {Fishman}, {Bhat},
  {Briggs}, {Koshut} et~al.}]{Kouveliotou1993}
{Kouveliotou} C, {Meegan} CA, {Fishman} GJ, {Bhat} NP, {Briggs} MS, {Koshut}
  TM, et~al.
\newblock {Identification of Two Classes of Gamma-Ray Bursts}.
\newblock {\em \apjl\/} {\bf 413} (1993) L101.
\newblock \doi{10.1086/186969}.

\bibitem[{{Norris} and {Bonnell}(2006)}]{Norris2006}
{Norris} JP, {Bonnell} JT.
\newblock {Short Gamma-Ray Bursts with Extended Emission}.
\newblock {\em \apj\/} {\bf 643} (2006) 266--275.
\newblock \doi{10.1086/502796}.

\bibitem[{{Norris} et~al.(2010){Norris}, {Gehrels}, and {Scargle}}]{Norris2010}
{Norris} JP, {Gehrels} N, {Scargle} JD.
\newblock {Threshold for Extended Emission in Short Gamma-ray Bursts}.
\newblock {\em \apj\/} {\bf 717} (2010) 411--419.
\newblock \doi{10.1088/0004-637X/717/1/411}.

\bibitem[{{Dainotti} et~al.(2010){Dainotti}, {Willingale}, {Capozziello},
  {Fabrizio Cardone}, and {Ostrowski}}]{Dainotti2010}
{Dainotti} MG, {Willingale} R, {Capozziello} S, {Fabrizio Cardone} V,
  {Ostrowski} M.
\newblock {Discovery of a Tight Correlation for Gamma-ray Burst Afterglows with
  ``Canonical'' Light Curves}.
\newblock {\em \apjl\/} {\bf 722} (2010) L215--L219.
\newblock \doi{10.1088/2041-8205/722/2/L215}.

\bibitem[{{Barkov} and {Pozanenko}(2011)}]{Barkov2011}
{Barkov} MV, {Pozanenko} AS.
\newblock {Model of the extended emission of short gamma-ray bursts}.
\newblock {\em \mnras\/} {\bf 417} (2011) 2161--2165.
\newblock \doi{10.1111/j.1365-2966.2011.19398.x}.

\bibitem[{{Dainotti} et~al.(2017){Dainotti}, {Hernandez}, {Postnikov},
  {Nagataki}, {O'brien}, {Willingale} et~al.}]{Dainotti2017}
{Dainotti} MG, {Hernandez} X, {Postnikov} S, {Nagataki} S, {O'brien} P,
  {Willingale} R, et~al.
\newblock {A Study of the Gamma-Ray Burst Fundamental Plane}.
\newblock {\em \apj\/} {\bf 848} (2017) 88.
\newblock \doi{10.3847/1538-4357/aa8a6b}.

\bibitem[{{L{\"u}} et~al.(2010){L{\"u}}, {Liang}, {Zhang}, and
  {Zhang}}]{Lu2010}
{L{\"u}} HJ, {Liang} EW, {Zhang} BB, {Zhang} B.
\newblock {A New Classification Method for Gamma-ray Bursts}.
\newblock {\em \apj\/} {\bf 725} (2010) 1965--1970.
\newblock \doi{10.1088/0004-637X/725/2/1965}.

\bibitem[{{Bromberg} et~al.(2013){Bromberg}, {Nakar}, {Piran}, and
  {Sari}}]{Bromberg2013}
{Bromberg} O, {Nakar} E, {Piran} T, {Sari} R.
\newblock {Short versus Long and Collapsars versus Non-collapsars: A
  Quantitative Classification of Gamma-Ray Bursts}.
\newblock {\em \apj\/} {\bf 764} (2013) 179.
\newblock \doi{10.1088/0004-637X/764/2/179}.

\bibitem[{{Woosley} and {Bloom}(2006)}]{Woosley2006}
{Woosley} SE, {Bloom} JS.
\newblock {The Supernova Gamma-Ray Burst Connection}.
\newblock {\em \araa\/} {\bf 44} (2006) 507--556.
\newblock \doi{10.1146/annurev.astro.43.072103.150558}.

\bibitem[{{Woosley}(1993)}]{Woosley1993}
{Woosley} SE.
\newblock {Gamma-Ray Bursts from Stellar Mass Accretion Disks around Black
  Holes}.
\newblock {\em \apj\/} {\bf 405} (1993) 273.
\newblock \doi{10.1086/172359}.

\bibitem[{{Woosley} and {Heger}(2006)}]{Woosley2006b}
{Woosley} SE, {Heger} A.
\newblock {The Progenitor Stars of Gamma-Ray Bursts}.
\newblock {\em \apj\/} {\bf 637} (2006) 914--921.
\newblock \doi{10.1086/498500}.

\bibitem[{{Yoon} et~al.(2006){Yoon}, {Langer}, and {Norman}}]{Yoon2006}
{Yoon} SC, {Langer} N, {Norman} C.
\newblock {Single star progenitors of long gamma-ray bursts. I. Model grids and
  redshift dependent GRB rate}.
\newblock {\em \aap\/} {\bf 460} (2006) 199--208.
\newblock \doi{10.1051/0004-6361:20065912}.

\bibitem[{{Bucciantini} et~al.(2008){Bucciantini}, {Quataert}, {Arons},
  {Metzger}, and {Thompson}}]{Bucciantini2008}
{Bucciantini} N, {Quataert} E, {Arons} J, {Metzger} BD, {Thompson} TA.
\newblock {Relativistic jets and long-duration gamma-ray bursts from the birth
  of magnetars}.
\newblock {\em \mnras\/} {\bf 383} (2008) L25--L29.
\newblock \doi{10.1111/j.1745-3933.2007.00403.x}.

\bibitem[{{Bucciantini} et~al.(2009){Bucciantini}, {Quataert}, {Metzger},
  {Thompson}, {Arons}, and {Del Zanna}}]{Bucciantini2009}
{Bucciantini} N, {Quataert} E, {Metzger} BD, {Thompson} TA, {Arons} J, {Del
  Zanna} L.
\newblock {Magnetized relativistic jets and long-duration GRBs from magnetar
  spin-down during core-collapse supernovae}.
\newblock {\em \mnras\/} {\bf 396} (2009) 2038--2050.
\newblock \doi{10.1111/j.1365-2966.2009.14940.x}.

\bibitem[{{Metzger} et~al.(2011){Metzger}, {Giannios}, {Thompson},
  {Bucciantini}, and {Quataert}}]{Metzger2011}
{Metzger} BD, {Giannios} D, {Thompson} TA, {Bucciantini} N, {Quataert} E.
\newblock {The protomagnetar model for gamma-ray bursts}.
\newblock {\em \mnras\/} {\bf 413} (2011) 2031--2056.
\newblock \doi{10.1111/j.1365-2966.2011.18280.x}.

\bibitem[{{Lee} et~al.(2000){Lee}, {Wijers}, and {Brown}}]{Lee2000}
{Lee} HK, {Wijers} RAMJ, {Brown} GE.
\newblock {The Blandford-Znajek process as a central engine for a gamma-ray
  burst}.
\newblock {\em \physrep\/} {\bf 325} (2000) 83--114.
\newblock \doi{10.1016/S0370-1573(99)00084-8}.

\bibitem[{{Lei} et~al.(2013){Lei}, {Zhang}, and {Liang}}]{Lei2013}
{Lei} WH, {Zhang} B, {Liang} EW.
\newblock {Hyperaccreting Black Hole as Gamma-Ray Burst Central Engine. I.
  Baryon Loading in Gamma-Ray Burst Jets}.
\newblock {\em \apj\/} {\bf 765} (2013) 125.
\newblock \doi{10.1088/0004-637X/765/2/125}.

\bibitem[{{Meegan} et~al.(2009){Meegan}, {Lichti}, {Bhat}, {Bissaldi},
  {Briggs}, {Connaughton} et~al.}]{Meegan2009}
{Meegan} C, {Lichti} G, {Bhat} PN, {Bissaldi} E, {Briggs} MS, {Connaughton} V,
  et~al.
\newblock {The Fermi Gamma-ray Burst Monitor}.
\newblock {\em \apj\/} {\bf 702} (2009) 791--804.
\newblock \doi{10.1088/0004-637X/702/1/791}.

\bibitem[{{Belczynski} et~al.(2006){Belczynski}, {Perna}, {Bulik}, {Kalogera},
  {Ivanova}, and {Lamb}}]{Belczynski2006}
{Belczynski} K, {Perna} R, {Bulik} T, {Kalogera} V, {Ivanova} N, {Lamb} DQ.
\newblock {A Study of Compact Object Mergers as Short Gamma-Ray Burst
  Progenitors}.
\newblock {\em \apj\/} {\bf 648} (2006) 1110--1116.
\newblock \doi{10.1086/505169}.

\bibitem[{{Lee} and {Ramirez-Ruiz}(2007)}]{Lee2007}
{Lee} WH, {Ramirez-Ruiz} E.
\newblock {The progenitors of short gamma-ray bursts}.
\newblock {\em New Journal of Physics\/} {\bf 9} (2007) 17.
\newblock \doi{10.1088/1367-2630/9/1/017}.

\bibitem[{{Fong} and {Berger}(2013)}]{Fong2013a}
{Fong} W, {Berger} E.
\newblock {The Locations of Short Gamma-Ray Bursts as Evidence for Compact
  Object Binary Progenitors}.
\newblock {\em \apj\/} {\bf 776} (2013) 18.
\newblock \doi{10.1088/0004-637X/776/1/18}.

\bibitem[{{Giacomazzo} et~al.(2013){Giacomazzo}, {Perna}, {Rezzolla}, {Troja},
  and {Lazzati}}]{Giacomazzo2013}
{Giacomazzo} B, {Perna} R, {Rezzolla} L, {Troja} E, {Lazzati} D.
\newblock {Compact Binary Progenitors of Short Gamma-Ray Bursts}.
\newblock {\em \apjl\/} {\bf 762} (2013) L18.
\newblock \doi{10.1088/2041-8205/762/2/L18}.

\bibitem[{{Ruiz} et~al.(2016){Ruiz}, {Lang}, {Paschalidis}, and
  {Shapiro}}]{Ruiz2016}
{Ruiz} M, {Lang} RN, {Paschalidis} V, {Shapiro} SL.
\newblock {Binary Neutron Star Mergers: A Jet Engine for Short Gamma-Ray
  Bursts}.
\newblock {\em \apjl\/} {\bf 824} (2016) L6.
\newblock \doi{10.3847/2041-8205/824/1/L6}.

\bibitem[{{Ciolfi}(2018)}]{Ciolfi2018}
{Ciolfi} R.
\newblock {Short gamma-ray burst central engines}.
\newblock {\em International Journal of Modern Physics D\/} {\bf 27} (2018)
  1842004.
\newblock \doi{10.1142/S021827181842004X}.

\bibitem[{{Nakar} et~al.(2006){Nakar}, {Gal-Yam}, and {Fox}}]{Nakar2006}
{Nakar} E, {Gal-Yam} A, {Fox} DB.
\newblock {The Local Rate and the Progenitor Lifetimes of Short-Hard Gamma-Ray
  Bursts: Synthesis and Predictions for the Laser Interferometer
  Gravitational-Wave Observatory}.
\newblock {\em \apj\/} {\bf 650} (2006) 281--290.
\newblock \doi{10.1086/505855}.

\bibitem[{{Kiuchi} et~al.(2010){Kiuchi}, {Sekiguchi}, {Shibata}, and
  {Taniguchi}}]{Kiuchi2010}
{Kiuchi} K, {Sekiguchi} Y, {Shibata} M, {Taniguchi} K.
\newblock {Exploring Binary-Neutron-Star-Merger Scenario of Short-Gamma-Ray
  Bursts by Gravitational-Wave Observation}.
\newblock {\em \prl\/} {\bf 104} (2010) 141101.
\newblock \doi{10.1103/PhysRevLett.104.141101}.

\bibitem[{{Schutz}(2011)}]{Schutz2011}
{Schutz} BF.
\newblock {Networks of gravitational wave detectors and three figures of
  merit}.
\newblock {\em Classical and Quantum Gravity\/} {\bf 28} (2011) 125023.
\newblock \doi{10.1088/0264-9381/28/12/125023}.

\bibitem[{{Perna} et~al.(2016){Perna}, {Lazzati}, and {Giacomazzo}}]{Perna2016}
{Perna} R, {Lazzati} D, {Giacomazzo} B.
\newblock {Short Gamma-Ray Bursts from the Merger of Two Black Holes}.
\newblock {\em \apjl\/} {\bf 821} (2016) L18.
\newblock \doi{10.3847/2041-8205/821/1/L18}.

\bibitem[{{Liu} et~al.(2016){Liu}, {Romero}, {Liu}, and {Li}}]{Liu2016}
{Liu} T, {Romero} GE, {Liu} ML, {Li} A.
\newblock {Fast Radio Bursts and Their Gamma-Ray or Radio Afterglows as
  Kerr-Newman Black Hole Binaries}.
\newblock {\em \apj\/} {\bf 826} (2016) 82.
\newblock \doi{10.3847/0004-637X/826/1/82}.

\bibitem[{{Zhang}(2016)}]{Zhang2016}
{Zhang} B.
\newblock {Mergers of Charged Black Holes: Gravitational-wave Events, Short
  Gamma-Ray Bursts, and Fast Radio Bursts}.
\newblock {\em \apjl\/} {\bf 827} (2016) L31.
\newblock \doi{10.3847/2041-8205/827/2/L31}.

\bibitem[{{Abbott} et~al.(2017{\natexlab{a}}){Abbott}, {Abbott}, {Abbott},
  {Acernese}, {Ackley}, {Adams} et~al.}]{Abbott2017a}
{Abbott} BP, {Abbott} R, {Abbott} TD, {Acernese} F, {Ackley} K, {Adams} C,
  et~al.
\newblock {GW170817: Observation of Gravitational Waves from a Binary Neutron
  Star Inspiral}.
\newblock {\em \prl\/} {\bf 119} (2017{\natexlab{a}}) 161101.
\newblock \doi{10.1103/PhysRevLett.119.161101}.

\bibitem[{{Abbott} et~al.(2017{\natexlab{b}}){Abbott}, {Abbott}, {Abbott},
  {Acernese}, {Ackley}, {Adams} et~al.}]{Abbott2017c}
{Abbott} BP, {Abbott} R, {Abbott} TD, {Acernese} F, {Ackley} K, {Adams} C,
  et~al.
\newblock {Gravitational Waves and Gamma-Rays from a Binary Neutron Star
  Merger: GW170817 and GRB 170817A}.
\newblock {\em \apjl\/} {\bf 848} (2017{\natexlab{b}}) L13.
\newblock \doi{10.3847/2041-8213/aa920c}.

\bibitem[{{Goldstein} et~al.(2017){Goldstein}, {Veres}, {Burns}, {Briggs},
  {Hamburg}, {Kocevski} et~al.}]{Goldstein2017}
{Goldstein} A, {Veres} P, {Burns} E, {Briggs} MS, {Hamburg} R, {Kocevski} D,
  et~al.
\newblock {An Ordinary Short Gamma-Ray Burst with Extraordinary Implications:
  Fermi-GBM Detection of GRB 170817A}.
\newblock {\em \apjl\/} {\bf 848} (2017) L14.
\newblock \doi{10.3847/2041-8213/aa8f41}.

\bibitem[{{Savchenko} et~al.(2017){Savchenko}, {Ferrigno}, {Kuulkers},
  {Bazzano}, {Bozzo}, {Brandt} et~al.}]{Savchenko2017}
{Savchenko} V, {Ferrigno} C, {Kuulkers} E, {Bazzano} A, {Bozzo} E, {Brandt} S,
  et~al.
\newblock {INTEGRAL Detection of the First Prompt Gamma-Ray Signal Coincident
  with the Gravitational-wave Event GW170817}.
\newblock {\em \apjl\/} {\bf 848} (2017) L15.
\newblock \doi{10.3847/2041-8213/aa8f94}.

\bibitem[{{Zhang} et~al.(2018){Zhang}, {Zhang}, {Sun}, {Lei}, {Gao}, {Li}
  et~al.}]{Zhang2018}
{Zhang} BB, {Zhang} B, {Sun} H, {Lei} WH, {Gao} H, {Li} Y, et~al.
\newblock {A peculiar low-luminosity short gamma-ray burst from a double
  neutron star merger progenitor}.
\newblock {\em Nature Communications\/} {\bf 9} (2018) 447.
\newblock \doi{10.1038/s41467-018-02847-3}.

\bibitem[{{Abbott} et~al.(2017{\natexlab{c}}){Abbott}, {Abbott}, {Abbott},
  {Acernese}, {Ackley}, {Adams} et~al.}]{Abbott2017b}
{Abbott} BP, {Abbott} R, {Abbott} TD, {Acernese} F, {Ackley} K, {Adams} C,
  et~al.
\newblock {Multi-messenger Observations of a Binary Neutron Star Merger}.
\newblock {\em \apjl\/} {\bf 848} (2017{\natexlab{c}}) L12.
\newblock \doi{10.3847/2041-8213/aa91c9}.

\bibitem[{{Coulter} et~al.(2017){Coulter}, {Foley}, {Kilpatrick}, {Drout},
  {Piro}, {Shappee} et~al.}]{Coulter2017}
{Coulter} DA, {Foley} RJ, {Kilpatrick} CD, {Drout} MR, {Piro} AL, {Shappee} BJ,
  et~al.
\newblock {Swope Supernova Survey 2017a (SSS17a), the optical counterpart to a
  gravitational wave source}.
\newblock {\em Science\/} {\bf 358} (2017) 1556--1558.
\newblock \doi{10.1126/science.aap9811}.

\bibitem[{{Cowperthwaite} et~al.(2017){Cowperthwaite}, {Berger}, {Villar},
  {Metzger}, {Nicholl}, {Chornock} et~al.}]{Cowperthwaite2017}
{Cowperthwaite} PS, {Berger} E, {Villar} VA, {Metzger} BD, {Nicholl} M,
  {Chornock} R, et~al.
\newblock {The Electromagnetic Counterpart of the Binary Neutron Star Merger
  LIGO/Virgo GW170817. II. UV, Optical, and Near-infrared Light Curves and
  Comparison to Kilonova Models}.
\newblock {\em \apjl\/} {\bf 848} (2017) L17.
\newblock \doi{10.3847/2041-8213/aa8fc7}.

\bibitem[{{Pian} et~al.(2017){Pian}, {D'Avanzo}, {Benetti}, {Branchesi},
  {Brocato}, {Campana} et~al.}]{Pian2017}
{Pian} E, {D'Avanzo} P, {Benetti} S, {Branchesi} M, {Brocato} E, {Campana} S,
  et~al.
\newblock {Spectroscopic identification of r-process nucleosynthesis in a
  double neutron-star merger}.
\newblock {\em \nat\/} {\bf 551} (2017) 67--70.
\newblock \doi{10.1038/nature24298}.

\bibitem[{{Kasen} et~al.(2017){Kasen}, {Metzger}, {Barnes}, {Quataert}, and
  {Ramirez-Ruiz}}]{Kasen2017}
{Kasen} D, {Metzger} B, {Barnes} J, {Quataert} E, {Ramirez-Ruiz} E.
\newblock {Origin of the heavy elements in binary neutron-star mergers from a
  gravitational-wave event}.
\newblock {\em \nat\/} {\bf 551} (2017) 80--84.
\newblock \doi{10.1038/nature24453}.

\bibitem[{{Troja} et~al.(2017){Troja}, {Piro}, {van Eerten}, {Wollaeger}, {Im},
  {Fox} et~al.}]{Troja2017}
{Troja} E, {Piro} L, {van Eerten} H, {Wollaeger} RT, {Im} M, {Fox} OD, et~al.
\newblock {The X-ray counterpart to the gravitational-wave event GW170817}.
\newblock {\em \nat\/} {\bf 551} (2017) 71--74.
\newblock \doi{10.1038/nature24290}.

\bibitem[{{Smartt} et~al.(2017){Smartt}, {Chen}, {Jerkstrand}, {Coughlin},
  {Kankare}, {Sim} et~al.}]{Smartt2017}
{Smartt} SJ, {Chen} TW, {Jerkstrand} A, {Coughlin} M, {Kankare} E, {Sim} SA,
  et~al.
\newblock {A kilonova as the electromagnetic counterpart to a
  gravitational-wave source}.
\newblock {\em \nat\/} {\bf 551} (2017) 75--79.
\newblock \doi{10.1038/nature24303}.

\bibitem[{{Tanvir} et~al.(2017){Tanvir}, {Levan}, {Gonz{\'a}lez-Fern{\'a}ndez},
  {Korobkin}, {Mandel}, {Rosswog} et~al.}]{Tanvir2017}
{Tanvir} NR, {Levan} AJ, {Gonz{\'a}lez-Fern{\'a}ndez} C, {Korobkin} O, {Mandel}
  I, {Rosswog} S, et~al.
\newblock {The Emergence of a Lanthanide-rich Kilonova Following the Merger of
  Two Neutron Stars}.
\newblock {\em \apjl\/} {\bf 848} (2017) L27.
\newblock \doi{10.3847/2041-8213/aa90b6}.

\bibitem[{{Hallinan} et~al.(2017){Hallinan}, {Corsi}, {Mooley}, {Hotokezaka},
  {Nakar}, {Kasliwal} et~al.}]{Hallinan2017}
{Hallinan} G, {Corsi} A, {Mooley} KP, {Hotokezaka} K, {Nakar} E, {Kasliwal} MM,
  et~al.
\newblock {A radio counterpart to a neutron star merger}.
\newblock {\em Science\/} {\bf 358} (2017) 1579--1583.
\newblock \doi{10.1126/science.aap9855}.

\bibitem[{{Soares-Santos} et~al.(2017){Soares-Santos}, {Holz}, {Annis},
  {Chornock}, {Herner}, {Berger} et~al.}]{Soares-Santos2017}
{Soares-Santos} M, {Holz} DE, {Annis} J, {Chornock} R, {Herner} K, {Berger} E,
  et~al.
\newblock {The Electromagnetic Counterpart of the Binary Neutron Star Merger
  LIGO/Virgo GW170817. I. Discovery of the Optical Counterpart Using the Dark
  Energy Camera}.
\newblock {\em \apjl\/} {\bf 848} (2017) L16.
\newblock \doi{10.3847/2041-8213/aa9059}.

\bibitem[{{Evans} et~al.(2017){Evans}, {Cenko}, {Kennea}, {Emery}, {Kuin},
  {Korobkin} et~al.}]{Evans2017}
{Evans} PA, {Cenko} SB, {Kennea} JA, {Emery} SWK, {Kuin} NPM, {Korobkin} O,
  et~al.
\newblock {Swift and NuSTAR observations of GW170817: Detection of a blue
  kilonova}.
\newblock {\em Science\/} {\bf 358} (2017) 1565--1570.
\newblock \doi{10.1126/science.aap9580}.

\bibitem[{{Arcavi} et~al.(2017){Arcavi}, {Hosseinzadeh}, {Howell}, {McCully},
  {Poznanski}, {Kasen} et~al.}]{Arcavi2017}
{Arcavi} I, {Hosseinzadeh} G, {Howell} DA, {McCully} C, {Poznanski} D, {Kasen}
  D, et~al.
\newblock {Optical emission from a kilonova following a
  gravitational-wave-detected neutron-star merger}.
\newblock {\em \nat\/} {\bf 551} (2017) 64--66.
\newblock \doi{10.1038/nature24291}.

\bibitem[{{Margutti} et~al.(2017){Margutti}, {Berger}, {Fong}, {Guidorzi},
  {Alexander}, {Metzger} et~al.}]{Margutti2017}
{Margutti} R, {Berger} E, {Fong} W, {Guidorzi} C, {Alexander} KD, {Metzger} BD,
  et~al.
\newblock {The Electromagnetic Counterpart of the Binary Neutron Star Merger
  LIGO/Virgo GW170817. V. Rising X-Ray Emission from an Off-axis Jet}.
\newblock {\em \apjl\/} {\bf 848} (2017) L20.
\newblock \doi{10.3847/2041-8213/aa9057}.

\bibitem[{{Mooley} et~al.(2018{\natexlab{a}}){Mooley}, {Nakar}, {Hotokezaka},
  {Hallinan}, {Corsi}, {Frail} et~al.}]{Mooley2018a}
{Mooley} KP, {Nakar} E, {Hotokezaka} K, {Hallinan} G, {Corsi} A, {Frail} DA,
  et~al.
\newblock {A mildly relativistic wide-angle outflow in the neutron-star merger
  event GW170817}.
\newblock {\em \nat\/} {\bf 554} (2018{\natexlab{a}}) 207--210.
\newblock \doi{10.1038/nature25452}.

\bibitem[{{Margutti} et~al.(2018){Margutti}, {Alexander}, {Xie}, {Sironi},
  {Metzger}, {Kathirgamaraju} et~al.}]{Margutti2018}
{Margutti} R, {Alexander} KD, {Xie} X, {Sironi} L, {Metzger} BD,
  {Kathirgamaraju} A, et~al.
\newblock {The Binary Neutron Star Event LIGO/Virgo GW170817 160 Days after
  Merger: Synchrotron Emission across the Electromagnetic Spectrum}.
\newblock {\em \apjl\/} {\bf 856} (2018) L18.
\newblock \doi{10.3847/2041-8213/aab2ad}.

\bibitem[{{Alexander} et~al.(2017){Alexander}, {Berger}, {Fong}, {Williams},
  {Guidorzi}, {Margutti} et~al.}]{Alexander2017}
{Alexander} KD, {Berger} E, {Fong} W, {Williams} PKG, {Guidorzi} C, {Margutti}
  R, et~al.
\newblock {The Electromagnetic Counterpart of the Binary Neutron Star Merger
  LIGO/Virgo GW170817. VI. Radio Constraints on a Relativistic Jet and
  Predictions for Late-time Emission from the Kilonova Ejecta}.
\newblock {\em \apjl\/} {\bf 848} (2017) L21.
\newblock \doi{10.3847/2041-8213/aa905d}.

\bibitem[{{Mooley} et~al.(2018{\natexlab{b}}){Mooley}, {Deller}, {Gottlieb},
  {Nakar}, {Hallinan}, {Bourke} et~al.}]{Mooley2018b}
{Mooley} KP, {Deller} AT, {Gottlieb} O, {Nakar} E, {Hallinan} G, {Bourke} S,
  et~al.
\newblock {Superluminal motion of a relativistic jet in the neutron-star merger
  GW170817}.
\newblock {\em \nat\/} {\bf 561} (2018{\natexlab{b}}) 355--359.
\newblock \doi{10.1038/s41586-018-0486-3}.

\bibitem[{{Valenti} et~al.(2017){Valenti}, {Sand}, {Yang}, {Cappellaro},
  {Tartaglia}, {Corsi} et~al.}]{Valenti2017}
{Valenti} S, {Sand} DJ, {Yang} S, {Cappellaro} E, {Tartaglia} L, {Corsi} A,
  et~al.
\newblock {The Discovery of the Electromagnetic Counterpart of GW170817:
  Kilonova AT 2017gfo/DLT17ck}.
\newblock {\em \apjl\/} {\bf 848} (2017) L24.
\newblock \doi{10.3847/2041-8213/aa8edf}.

\bibitem[{{Haggard} et~al.(2017){Haggard}, {Nynka}, {Ruan}, {Kalogera},
  {Cenko}, {Evans} et~al.}]{Haggard2017}
{Haggard} D, {Nynka} M, {Ruan} JJ, {Kalogera} V, {Cenko} SB, {Evans} P, et~al.
\newblock {A Deep Chandra X-Ray Study of Neutron Star Coalescence GW170817}.
\newblock {\em \apjl\/} {\bf 848} (2017) L25.
\newblock \doi{10.3847/2041-8213/aa8ede}.

\bibitem[{{Lyman} et~al.(2018){Lyman}, {Lamb}, {Levan}, {Mandel}, {Tanvir},
  {Kobayashi} et~al.}]{Lyman2018}
{Lyman} JD, {Lamb} GP, {Levan} AJ, {Mandel} I, {Tanvir} NR, {Kobayashi} S,
  et~al.
\newblock {The optical afterglow of the short gamma-ray burst associated with
  GW170817}.
\newblock {\em Nature Astronomy\/} {\bf 2} (2018) 751--754.
\newblock \doi{10.1038/s41550-018-0511-3}.

\bibitem[{{Ghirlanda} et~al.(2019){Ghirlanda}, {Salafia}, {Paragi},
  {Giroletti}, {Yang}, {Marcote} et~al.}]{Ghirlanda2019}
{Ghirlanda} G, {Salafia} OS, {Paragi} Z, {Giroletti} M, {Yang} J, {Marcote} B,
  et~al.
\newblock {Compact radio emission indicates a structured jet was produced by a
  binary neutron star merger}.
\newblock {\em Science\/} {\bf 363} (2019) 968--971.
\newblock \doi{10.1126/science.aau8815}.

\bibitem[{{D'Avanzo} et~al.(2018){D'Avanzo}, {Campana}, {Salafia}, {Ghirland
  a}, {Ghisellini}, {Melandri} et~al.}]{Davanzo2018}
{D'Avanzo} P, {Campana} S, {Salafia} OS, {Ghirland a} G, {Ghisellini} G,
  {Melandri} A, et~al.
\newblock {The evolution of the X-ray afterglow emission of GW 170817/ GRB
  170817A in XMM-Newton observations}.
\newblock {\em \aap\/} {\bf 613} (2018) L1.
\newblock \doi{10.1051/0004-6361/201832664}.

\bibitem[{{Resmi} et~al.(2018){Resmi}, {Schulze}, {Ishwara-Chandra}, {Misra},
  {Buchner}, {De Pasquale} et~al.}]{Resmi2018}
{Resmi} L, {Schulze} S, {Ishwara-Chandra} CH, {Misra} K, {Buchner} J, {De
  Pasquale} M, et~al.
\newblock {Low-frequency View of GW170817/GRB 170817A with the Giant Metrewave
  Radio Telescope}.
\newblock {\em \apj\/} {\bf 867} (2018) 57.
\newblock \doi{10.3847/1538-4357/aae1a6}.

\bibitem[{{Lamb} et~al.(2019{\natexlab{a}}){Lamb}, {Lyman}, {Levan}, {Tanvir},
  {Kangas}, {Fruchter} et~al.}]{Lamb2019}
{Lamb} GP, {Lyman} JD, {Levan} AJ, {Tanvir} NR, {Kangas} T, {Fruchter} AS,
  et~al.
\newblock {The Optical Afterglow of GW170817 at One Year Post-merger}.
\newblock {\em \apjl\/} {\bf 870} (2019{\natexlab{a}}) L15.
\newblock \doi{10.3847/2041-8213/aaf96b}.

\bibitem[{{Kim} et~al.(2017){Kim}, {Schulze}, {Resmi},
  {Gonz{\'a}lez-L{\'o}pez}, {Higgins}, {Ishwara-Chand ra} et~al.}]{Kim2017}
{Kim} S, {Schulze} S, {Resmi} L, {Gonz{\'a}lez-L{\'o}pez} J, {Higgins} AB,
  {Ishwara-Chand ra} CH, et~al.
\newblock {ALMA and GMRT Constraints on the Off-axis Gamma-Ray Burst 170817A
  from the Binary Neutron Star Merger GW170817}.
\newblock {\em \apjl\/} {\bf 850} (2017) L21.
\newblock \doi{10.3847/2041-8213/aa970b}.

\bibitem[{{Abdalla} et~al.(2017){Abdalla}, {Abramowski}, {Aharonian}, {Ait
  Benkhali}, {Ang{\"u}ner}, {Arakawa} et~al.}]{Abdalla2017}
{Abdalla} H, {Abramowski} A, {Aharonian} F, {Ait Benkhali} F, {Ang{\"u}ner} EO,
  {Arakawa} M, et~al.
\newblock {TeV Gamma-Ray Observations of the Binary Neutron Star Merger
  GW170817 with H.E.S.S.}
\newblock {\em \apjl\/} {\bf 850} (2017) L22.
\newblock \doi{10.3847/2041-8213/aa97d2}.

\bibitem[{{Siebert} et~al.(2017){Siebert}, {Foley}, {Drout}, {Kilpatrick},
  {Shappee}, {Coulter} et~al.}]{Siebert2017}
{Siebert} MR, {Foley} RJ, {Drout} MR, {Kilpatrick} CD, {Shappee} BJ, {Coulter}
  DA, et~al.
\newblock {The Unprecedented Properties of the First Electromagnetic
  Counterpart to a Gravitational-wave Source}.
\newblock {\em \apjl\/} {\bf 848} (2017) L26.
\newblock \doi{10.3847/2041-8213/aa905e}.

\bibitem[{{Villar} et~al.(2018){Villar}, {Cowperthwaite}, {Berger},
  {Blanchard}, {Gomez}, {Alexander} et~al.}]{Villar2018}
{Villar} VA, {Cowperthwaite} PS, {Berger} E, {Blanchard} PK, {Gomez} S,
  {Alexander} KD, et~al.
\newblock {Spitzer Space Telescope Infrared Observations of the Binary Neutron
  Star Merger GW170817}.
\newblock {\em \apjl\/} {\bf 862} (2018) L11.
\newblock \doi{10.3847/2041-8213/aad281}.

\bibitem[{{Coughlin} and {Dietrich}(2019)}]{Coughlin2019}
{Coughlin} MW, {Dietrich} T.
\newblock {Can a black hole-neutron star merger explain GW170817, AT2017gfo,
  and GRB170817A?}
\newblock {\em \prd\/} {\bf 100} (2019) 043011.
\newblock \doi{10.1103/PhysRevD.100.043011}.

\bibitem[{{Granot} et~al.(2017){Granot}, {Guetta}, and {Gill}}]{Granot2017}
{Granot} J, {Guetta} D, {Gill} R.
\newblock {Lessons from the Short GRB 170817A: The First Gravitational-wave
  Detection of a Binary Neutron Star Merger}.
\newblock {\em \apjl\/} {\bf 850} (2017) L24.
\newblock \doi{10.3847/2041-8213/aa991d}.

\bibitem[{{Lin} et~al.(2018){Lin}, {Liu}, {Lin}, {Wang}, {Gu}, and
  {Liang}}]{Lin2018}
{Lin} DB, {Liu} T, {Lin} J, {Wang} XG, {Gu} WM, {Liang} EW.
\newblock {First Electromagnetic Pulse Associated with a Gravitational-wave
  Event: Profile, Duration, and Delay}.
\newblock {\em \apj\/} {\bf 856} (2018) 90.
\newblock \doi{10.3847/1538-4357/aab3d7}.

\bibitem[{{Zhang}(2019)}]{Zhang2019a}
{Zhang} B.
\newblock {The delay time of gravitational wave {\textemdash} gamma-ray burst
  associations}.
\newblock {\em Frontiers of Physics\/} {\bf 14} (2019) 64402.
\newblock \doi{10.1007/s11467-019-0913-4}.

\bibitem[{{Lazzati} et~al.(2020){Lazzati}, {Ciolfi}, and {Perna}}]{Lazzati2020}
{Lazzati} D, {Ciolfi} R, {Perna} R.
\newblock {Intrinsic properties of the engine and jet that powered the short
  gamma-ray burst associated with GW170817}.
\newblock {\em arXiv e-prints\/}  (2020) arXiv:2004.10210.

\bibitem[{{Lucca} and {Sagunski}(2020)}]{Lucca2020}
{Lucca} M, {Sagunski} L.
\newblock {The lifetime of binary neutron star merger remnants}.
\newblock {\em Journal of High Energy Astrophysics\/} {\bf 27} (2020) 33--37.
\newblock \doi{10.1016/j.jheap.2020.04.003}.

\bibitem[{{Cook} et~al.(1994){Cook}, {Shapiro}, and {Teukolsky}}]{Cook1994}
{Cook} GB, {Shapiro} SL, {Teukolsky} SA.
\newblock {Rapidly Rotating Polytropes in General Relativity}.
\newblock {\em \apj\/} {\bf 422} (1994) 227.
\newblock \doi{10.1086/173721}.

\bibitem[{{Lasota} et~al.(1996){Lasota}, {Haensel}, and
  {Abramowicz}}]{Lasota1996}
{Lasota} JP, {Haensel} P, {Abramowicz} MA.
\newblock {Fast Rotation of Neutron Stars}.
\newblock {\em \apj\/} {\bf 456} (1996) 300.
\newblock \doi{10.1086/176650}.

\bibitem[{{Vietri} and {Stella}(1998)}]{Vietri1998}
{Vietri} M, {Stella} L.
\newblock {A Gamma-Ray Burst Model with Small Baryon Contamination}.
\newblock {\em \apjl\/} {\bf 507} (1998) L45--L48.
\newblock \doi{10.1086/311674}.

\bibitem[{{Ciolfi} and {Siegel}(2015)}]{Ciolfi2015}
{Ciolfi} R, {Siegel} DM.
\newblock {Short Gamma-Ray Bursts in the ``Time-reversal'' Scenario}.
\newblock {\em \apjl\/} {\bf 798} (2015) L36.
\newblock \doi{10.1088/2041-8205/798/2/L36}.

\bibitem[{{Matzner}(2003)}]{Matzner2003}
{Matzner} CD.
\newblock {Supernova hosts for gamma-ray burst jets: dynamical constraints}.
\newblock {\em \mnras\/} {\bf 345} (2003) 575--589.
\newblock \doi{10.1046/j.1365-8711.2003.06969.x}.

\bibitem[{{Morsony} et~al.(2007){Morsony}, {Lazzati}, and
  {Begelman}}]{Morsony2007}
{Morsony} BJ, {Lazzati} D, {Begelman} MC.
\newblock {Temporal and Angular Properties of Gamma-Ray Burst Jets Emerging
  from Massive Stars}.
\newblock {\em \apj\/} {\bf 665} (2007) 569--598.
\newblock \doi{10.1086/519483}.

\bibitem[{{Bromberg} et~al.(2011){Bromberg}, {Nakar}, {Piran}, and
  {Sari}}]{Bromberg2011}
{Bromberg} O, {Nakar} E, {Piran} T, {Sari} R.
\newblock {The Propagation of Relativistic Jets in External Media}.
\newblock {\em \apj\/} {\bf 740} (2011) 100.
\newblock \doi{10.1088/0004-637X/740/2/100}.

\bibitem[{{Lazzati} and {Perna}(2019)}]{Lazzati2019}
{Lazzati} D, {Perna} R.
\newblock {Jet-Cocoon Outflows from Neutron Star Mergers: Structure, Light
  Curves, and Fundamental Physics}.
\newblock {\em \apj\/} {\bf 881} (2019) 89.
\newblock \doi{10.3847/1538-4357/ab2e06}.

\bibitem[{{Ramirez-Ruiz} et~al.(2002){Ramirez-Ruiz}, {Celotti}, and
  {Rees}}]{Ramirez-Ruiz2002}
{Ramirez-Ruiz} E, {Celotti} A, {Rees} MJ.
\newblock {Events in the life of a cocoon surrounding a light, collapsar jet}.
\newblock {\em \mnras\/} {\bf 337} (2002) 1349--1356.
\newblock \doi{10.1046/j.1365-8711.2002.05995.x}.

\bibitem[{{Lazzati} et~al.(2017{\natexlab{a}}){Lazzati}, {Deich}, {Morsony},
  and {Workman}}]{Lazzati2017a}
{Lazzati} D, {Deich} A, {Morsony} BJ, {Workman} JC.
\newblock {Off-axis emission of short {\ensuremath{\gamma}}-ray bursts and the
  detectability of electromagnetic counterparts of gravitational-wave-detected
  binary mergers}.
\newblock {\em \mnras\/} {\bf 471} (2017{\natexlab{a}}) 1652--1661.
\newblock \doi{10.1093/mnras/stx1683}.

\bibitem[{{Lazzati} et~al.(2017{\natexlab{b}}){Lazzati},
  {L{\'o}pez-C{\'a}mara}, {Cantiello}, {Morsony}, {Perna}, and
  {Workman}}]{Lazzati2017b}
{Lazzati} D, {L{\'o}pez-C{\'a}mara} D, {Cantiello} M, {Morsony} BJ, {Perna} R,
  {Workman} JC.
\newblock {Off-axis Prompt X-Ray Transients from the Cocoon of Short Gamma-Ray
  Bursts}.
\newblock {\em \apjl\/} {\bf 848} (2017{\natexlab{b}}) L6.
\newblock \doi{10.3847/2041-8213/aa8f3d}.

\bibitem[{{Gill} et~al.(2019{\natexlab{a}}){Gill}, {Nathanail}, and
  {Rezzolla}}]{Gill2019}
{Gill} R, {Nathanail} A, {Rezzolla} L.
\newblock {When Did the Remnant of GW170817 Collapse to a Black Hole?}
\newblock {\em \apj\/} {\bf 876} (2019{\natexlab{a}}) 139.
\newblock \doi{10.3847/1538-4357/ab16da}.

\bibitem[{{Shoemaker} and {Murase}(2018)}]{Shoemaker2018}
{Shoemaker} IM, {Murase} K.
\newblock {Constraints from the time lag between gravitational waves and gamma
  rays: Implications of GW170817 and GRB 170817A}.
\newblock {\em \prd\/} {\bf 97} (2018) 083013.
\newblock \doi{10.1103/PhysRevD.97.083013}.

\bibitem[{{Hamidani} et~al.(2020){Hamidani}, {Kiuchi}, and
  {Ioka}}]{Hamidani2020}
{Hamidani} H, {Kiuchi} K, {Ioka} K.
\newblock {Jet propagation in neutron star mergers and GW170817}.
\newblock {\em \mnras\/} {\bf 491} (2020) 3192--3216.
\newblock \doi{10.1093/mnras/stz3231}.

\bibitem[{{Beniamini} et~al.(2020{\natexlab{a}}){Beniamini}, {Duran},
  {Petropoulou}, and {Giannios}}]{Beniamini2020}
{Beniamini} P, {Duran} RB, {Petropoulou} M, {Giannios} D.
\newblock {Ready, Set, Launch: Time Interval between a Binary Neutron Star
  Merger and Short Gamma-Ray Burst Jet Formation}.
\newblock {\em \apjl\/} {\bf 895} (2020{\natexlab{a}}) L33.
\newblock \doi{10.3847/2041-8213/ab9223}.

\bibitem[{{Beniamini} et~al.(2020{\natexlab{b}}){Beniamini}, {Granot}, and
  {Gill}}]{Beniamini2020aa}
{Beniamini} P, {Granot} J, {Gill} R.
\newblock {Afterglow light curves from misaligned structured jets}.
\newblock {\em \mnras\/} {\bf 493} (2020{\natexlab{b}}) 3521--3534.
\newblock \doi{10.1093/mnras/staa538}.

\bibitem[{{Aloy} et~al.(2005){Aloy}, {Janka}, and {M{\"u}ller}}]{Aloy2005}
{Aloy} MA, {Janka} HT, {M{\"u}ller} E.
\newblock {Relativistic outflows from remnants of compact object mergers and
  their viability for short gamma-ray bursts}.
\newblock {\em \aap\/} {\bf 436} (2005) 273--311.
\newblock \doi{10.1051/0004-6361:20041865}.

\bibitem[{{Kathirgamaraju} et~al.(2019){Kathirgamaraju}, {Tchekhovskoy},
  {Giannios}, and {Barniol Duran}}]{Kathirgamaraju2019}
{Kathirgamaraju} A, {Tchekhovskoy} A, {Giannios} D, {Barniol Duran} R.
\newblock {EM counterparts of structured jets from 3D GRMHD simulations}.
\newblock {\em \mnras\/} {\bf 484} (2019) L98--L103.
\newblock \doi{10.1093/mnrasl/slz012}.

\bibitem[{{Salafia} et~al.(2020){Salafia}, {Barbieri}, {Ascenzi}, and
  {Toffano}}]{Salafia2020}
{Salafia} OS, {Barbieri} C, {Ascenzi} S, {Toffano} M.
\newblock {Gamma-ray burst jet propagation, development of angular structure,
  and the luminosity function}.
\newblock {\em \aap\/} {\bf 636} (2020) A105.
\newblock \doi{10.1051/0004-6361/201936335}.

\bibitem[{{Nagakura} et~al.(2014){Nagakura}, {Hotokezaka}, {Sekiguchi},
  {Shibata}, and {Ioka}}]{Nagakura2014}
{Nagakura} H, {Hotokezaka} K, {Sekiguchi} Y, {Shibata} M, {Ioka} K.
\newblock {Jet Collimation in the Ejecta of Double Neutron Star Mergers: A New
  Canonical Picture of Short Gamma-Ray Bursts}.
\newblock {\em \apjl\/} {\bf 784} (2014) L28.
\newblock \doi{10.1088/2041-8205/784/2/L28}.

\bibitem[{{Murguia-Berthier} et~al.(2014){Murguia-Berthier}, {Montes},
  {Ramirez-Ruiz}, {De Colle}, and {Lee}}]{Murguia-Berthier2014}
{Murguia-Berthier} A, {Montes} G, {Ramirez-Ruiz} E, {De Colle} F, {Lee} WH.
\newblock {Necessary Conditions for Short Gamma-Ray Burst Production in Binary
  Neutron Star Mergers}.
\newblock {\em \apjl\/} {\bf 788} (2014) L8.
\newblock \doi{10.1088/2041-8205/788/1/L8}.

\bibitem[{{Murguia-Berthier} et~al.(2017{\natexlab{a}}){Murguia-Berthier},
  {Ramirez-Ruiz}, {Montes}, {De Colle}, {Rezzolla}, {Rosswog}
  et~al.}]{Murguia-Berthier2017}
{Murguia-Berthier} A, {Ramirez-Ruiz} E, {Montes} G, {De Colle} F, {Rezzolla} L,
  {Rosswog} S, et~al.
\newblock {The Properties of Short Gamma-Ray Burst Jets Triggered by Neutron
  Star Mergers}.
\newblock {\em \apjl\/} {\bf 835} (2017{\natexlab{a}}) L34.
\newblock \doi{10.3847/2041-8213/aa5b9e}.

\bibitem[{{Murguia-Berthier} et~al.(2017{\natexlab{b}}){Murguia-Berthier},
  {Ramirez-Ruiz}, {Kilpatrick}, {Foley}, {Kasen}, {Lee}
  et~al.}]{Murguia-Berthier2017a}
{Murguia-Berthier} A, {Ramirez-Ruiz} E, {Kilpatrick} CD, {Foley} RJ, {Kasen} D,
  {Lee} WH, et~al.
\newblock {A Neutron Star Binary Merger Model for GW170817/GRB 170817A/SSS17a}.
\newblock {\em \apjl\/} {\bf 848} (2017{\natexlab{b}}) L34.
\newblock \doi{10.3847/2041-8213/aa91b3}.

\bibitem[{{Xie} et~al.(2018){Xie}, {Zrake}, and {MacFadyen}}]{Xie2018}
{Xie} X, {Zrake} J, {MacFadyen} A.
\newblock {Numerical Simulations of the Jet Dynamics and Synchrotron Radiation
  of Binary Neutron Star Merger Event GW170817/GRB 170817A}.
\newblock {\em \apj\/} {\bf 863} (2018) 58.
\newblock \doi{10.3847/1538-4357/aacf9c}.

\bibitem[{{Wu} and {MacFadyen}(2018)}]{Wu2018}
{Wu} Y, {MacFadyen} A.
\newblock {Constraining the Outflow Structure of the Binary Neutron Star Merger
  Event GW170817/GRB170817A with a Markov Chain Monte Carlo Analysis}.
\newblock {\em \apj\/} {\bf 869} (2018) 55.
\newblock \doi{10.3847/1538-4357/aae9de}.

\bibitem[{{Duffell} et~al.(2018){Duffell}, {Quataert}, {Kasen}, and
  {Klion}}]{Duffell2018}
{Duffell} PC, {Quataert} E, {Kasen} D, {Klion} H.
\newblock {Jet Dynamics in Compact Object Mergers: GW170817 Likely Had a
  Successful Jet}.
\newblock {\em \apj\/} {\bf 866} (2018) 3.
\newblock \doi{10.3847/1538-4357/aae084}.

\bibitem[{{Granot} et~al.(2018{\natexlab{a}}){Granot}, {De Colle}, and
  {Ramirez-Ruiz}}]{Granot2018a}
{Granot} J, {De Colle} F, {Ramirez-Ruiz} E.
\newblock {Off-axis afterglow light curves and images from 2D hydrodynamic
  simulations of double-sided GRB jets in a stratified external medium}.
\newblock {\em \mnras\/} {\bf 481} (2018{\natexlab{a}}) 2711--2720.
\newblock \doi{10.1093/mnras/sty2454}.

\bibitem[{{Geng} et~al.(2019){Geng}, {Zhang}, {K{\"o}lligan}, {Kuiper}, and
  {Huang}}]{Geng2019}
{Geng} JJ, {Zhang} B, {K{\"o}lligan} A, {Kuiper} R, {Huang} YF.
\newblock {Propagation of a Short GRB Jet in the Ejecta: Jet Launching Delay
  Time, Jet Structure, and GW170817/GRB 170817A}.
\newblock {\em \apjl\/} {\bf 877} (2019) L40.
\newblock \doi{10.3847/2041-8213/ab224b}.

\bibitem[{{Gill} et~al.(2019{\natexlab{b}}){Gill}, {Granot}, {De Colle}, and
  {Urrutia}}]{Gill2019c}
{Gill} R, {Granot} J, {De Colle} F, {Urrutia} G.
\newblock {Numerical Simulations of an Initially Top-hat Jet and the Afterglow
  of GW170817/GRB170817A}.
\newblock {\em \apj\/} {\bf 883} (2019{\natexlab{b}}) 15.
\newblock \doi{10.3847/1538-4357/ab3577}.

\bibitem[{{Hamidani} and {Ioka}(2020)}]{Hamidani2020a}
{Hamidani} H, {Ioka} K.
\newblock {Jet Propagation in Expanding Medium for Gamma-Ray Bursts}.
\newblock {\em arXiv e-prints\/}  (2020) arXiv:2007.10690.

\bibitem[{{Murguia-Berthier} et~al.(2020){Murguia-Berthier}, {Ramirez-Ruiz},
  {De Colle}, {Janiuk}, {Rosswog}, and {Lee}}]{Murguia2020}
{Murguia-Berthier} A, {Ramirez-Ruiz} E, {De Colle} F, {Janiuk} A, {Rosswog} S,
  {Lee} WH.
\newblock {The Fate of the Merger Remnant in GW170817 and its Imprint on the
  Jet Structure}.
\newblock {\em arXiv e-prints\/}  (2020) arXiv:2007.12245.

\bibitem[{{Takahashi} and {Ioka}(2020{\natexlab{a}})}]{Takahashi2020}
{Takahashi} K, {Ioka} K.
\newblock {Inverse reconstruction of jet structure from off-axis gamma-ray
  burst afterglows}.
\newblock {\em \mnras\/} {\bf 497} (2020{\natexlab{a}}) 1217--1235.
\newblock \doi{10.1093/mnras/staa1984}.

\bibitem[{{Takahashi} and {Ioka}(2020{\natexlab{b}})}]{Takahashi2020a}
{Takahashi} K, {Ioka} K.
\newblock {Diverse Jet Structures Consistent with the Off-axis Afterglow of GRB
  170817A}.
\newblock {\em arXiv e-prints\/}  (2020{\natexlab{b}}) arXiv:2007.13116.

\bibitem[{{Fong} et~al.(2015){Fong}, {Berger}, {Margutti}, and
  {Zauderer}}]{Fong2015}
{Fong} W, {Berger} E, {Margutti} R, {Zauderer} BA.
\newblock {A Decade of Short-duration Gamma-Ray Burst Broadband Afterglows:
  Energetics, Circumburst Densities, and Jet Opening Angles}.
\newblock {\em \apj\/} {\bf 815} (2015) 102.
\newblock \doi{10.1088/0004-637X/815/2/102}.

\bibitem[{{Metzger} and {Berger}(2012)}]{Metzger2012}
{Metzger} BD, {Berger} E.
\newblock {What is the Most Promising Electromagnetic Counterpart of a Neutron
  Star Binary Merger?}
\newblock {\em \apj\/} {\bf 746} (2012) 48.
\newblock \doi{10.1088/0004-637X/746/1/48}.

\bibitem[{{Ghirlanda} et~al.(2016){Ghirlanda}, {Salafia}, {Pescalli},
  {Ghisellini}, {Salvaterra}, {Chassande-Mottin} et~al.}]{Ghirlanda2016}
{Ghirlanda} G, {Salafia} OS, {Pescalli} A, {Ghisellini} G, {Salvaterra} R,
  {Chassande-Mottin} E, et~al.
\newblock {Short gamma-ray bursts at the dawn of the gravitational wave era}.
\newblock {\em \aap\/} {\bf 594} (2016) A84.
\newblock \doi{10.1051/0004-6361/201628993}.

\bibitem[{{Yamazaki} et~al.(2002){Yamazaki}, {Ioka}, and
  {Nakamura}}]{Yamazaki2002}
{Yamazaki} R, {Ioka} K, {Nakamura} T.
\newblock {X-Ray Flashes from Off-Axis Gamma-Ray Bursts}.
\newblock {\em \apjl\/} {\bf 571} (2002) L31--L35.
\newblock \doi{10.1086/341225}.

\bibitem[{{Yamazaki} et~al.(2003){Yamazaki}, {Yonetoku}, and
  {Nakamura}}]{Yamazaki2003}
{Yamazaki} R, {Yonetoku} D, {Nakamura} T.
\newblock {An Off-Axis Jet Model For GRB 980425 and Low-Energy Gamma-Ray
  Bursts}.
\newblock {\em \apjl\/} {\bf 594} (2003) L79--L82.
\newblock \doi{10.1086/378736}.

\bibitem[{{Lazzati} et~al.(2009){Lazzati}, {Morsony}, and
  {Begelman}}]{Lazzati2009}
{Lazzati} D, {Morsony} BJ, {Begelman} MC.
\newblock {Very High Efficiency Photospheric Emission in Long-Duration
  {\ensuremath{\gamma}}-Ray Bursts}.
\newblock {\em \apjl\/} {\bf 700} (2009) L47--L50.
\newblock \doi{10.1088/0004-637X/700/1/L47}.

\bibitem[{{Parsotan} et~al.(2018){Parsotan}, {L{\'o}pez-C{\'a}mara}, and
  {Lazzati}}]{Parsotan2018}
{Parsotan} T, {L{\'o}pez-C{\'a}mara} D, {Lazzati} D.
\newblock {Photospheric Emission from Variable Engine Gamma-Ray Burst
  Simulations}.
\newblock {\em \apj\/} {\bf 869} (2018) 103.
\newblock \doi{10.3847/1538-4357/aaeed1}.

\bibitem[{{Veres} et~al.(2018){Veres}, {M{\'e}sz{\'a}ros}, {Goldstein},
  {Fraija}, {Connaughton}, {Burns} et~al.}]{Veres2018}
{Veres} P, {M{\'e}sz{\'a}ros} P, {Goldstein} A, {Fraija} N, {Connaughton} V,
  {Burns} E, et~al.
\newblock {Gamma-ray burst models in light of the GRB 170817A - GW170817
  connection}.
\newblock {\em arXiv e-prints\/}  (2018) arXiv:1802.07328.

\bibitem[{{Kasliwal} et~al.(2017){Kasliwal}, {Nakar}, {Singer}, {Kaplan},
  {Cook}, {Van Sistine} et~al.}]{Kasliwal2017}
{Kasliwal} MM, {Nakar} E, {Singer} LP, {Kaplan} DL, {Cook} DO, {Van Sistine} A,
  et~al.
\newblock {Illuminating gravitational waves: A concordant picture of photons
  from a neutron star merger}.
\newblock {\em Science\/} {\bf 358} (2017) 1559--1565.
\newblock \doi{10.1126/science.aap9455}.

\bibitem[{{Gottlieb} et~al.(2018){Gottlieb}, {Nakar}, {Piran}, and
  {Hotokezaka}}]{Gottlieb2018}
{Gottlieb} O, {Nakar} E, {Piran} T, {Hotokezaka} K.
\newblock {A cocoon shock breakout as the origin of the
  {\ensuremath{\gamma}}-ray emission in GW170817}.
\newblock {\em \mnras\/} {\bf 479} (2018) 588--600.
\newblock \doi{10.1093/mnras/sty1462}.

\bibitem[{{Bromberg} et~al.(2018){Bromberg}, {Tchekhovskoy}, {Gottlieb},
  {Nakar}, and {Piran}}]{Bromberg2018}
{Bromberg} O, {Tchekhovskoy} A, {Gottlieb} O, {Nakar} E, {Piran} T.
\newblock {The {\ensuremath{\gamma}}-rays that accompanied GW170817 and the
  observational signature of a magnetic jet breaking out of NS merger ejecta}.
\newblock {\em \mnras\/} {\bf 475} (2018) 2971--2977.
\newblock \doi{10.1093/mnras/stx3316}.

\bibitem[{{Nakar} et~al.(2018){Nakar}, {Gottlieb}, {Piran}, {Kasliwal}, and
  {Hallinan}}]{Nakar2018a}
{Nakar} E, {Gottlieb} O, {Piran} T, {Kasliwal} MM, {Hallinan} G.
\newblock {From {\ensuremath{\gamma}} to Radio: The Electromagnetic Counterpart
  of GW170817}.
\newblock {\em \apj\/} {\bf 867} (2018) 18.
\newblock \doi{10.3847/1538-4357/aae205}.

\bibitem[{{Nakar} and {Sari}(2012)}]{Nakar2012x}
{Nakar} E, {Sari} R.
\newblock {Relativistic Shock Breakouts{\textemdash}A Variety of Gamma-Ray
  Flares: From Low-luminosity Gamma-Ray Bursts to Type Ia Supernovae}.
\newblock {\em \apj\/} {\bf 747} (2012) 88.
\newblock \doi{10.1088/0004-637X/747/2/88}.

\bibitem[{{Pozanenko} et~al.(2018){Pozanenko}, {Barkov}, {Minaev}, {Volnova},
  {Mazaeva}, {Moskvitin} et~al.}]{Pozanenko2018}
{Pozanenko} AS, {Barkov} MV, {Minaev} PY, {Volnova} AA, {Mazaeva} ED,
  {Moskvitin} AS, et~al.
\newblock {GRB 170817A Associated with GW170817: Multi-frequency Observations
  and Modeling of Prompt Gamma-Ray Emission}.
\newblock {\em \apjl\/} {\bf 852} (2018) L30.
\newblock \doi{10.3847/2041-8213/aaa2f6}.

\bibitem[{{Matsumoto} et~al.(2019){Matsumoto}, {Nakar}, and
  {Piran}}]{Matsumoto2019}
{Matsumoto} T, {Nakar} E, {Piran} T.
\newblock {Constraints on the emitting region of the gamma-rays observed in
  GW170817}.
\newblock {\em \mnras\/} {\bf 483} (2019) 1247--1255.
\newblock \doi{10.1093/mnras/sty3200}.

\bibitem[{{Ioka} and {Nakamura}(2019)}]{Ioka2019}
{Ioka} K, {Nakamura} T.
\newblock {Spectral puzzle of the off-axis gamma-ray burst in GW170817}.
\newblock {\em \mnras\/} {\bf 487} (2019) 4884--4889.
\newblock \doi{10.1093/mnras/stz1650}.

\bibitem[{{Meng} et~al.(2018){Meng}, {Geng}, {Zhang}, {Wei}, {Xiao}, {Liu}
  et~al.}]{Meng2018}
{Meng} YZ, {Geng} JJ, {Zhang} BB, {Wei} JJ, {Xiao} D, {Liu} LD, et~al.
\newblock {The Origin of the Prompt Emission for Short GRB 170817A: Photosphere
  Emission or Synchrotron Emission?}
\newblock {\em \apj\/} {\bf 860} (2018) 72.
\newblock \doi{10.3847/1538-4357/aac2d9}.

\bibitem[{{Kisaka} et~al.(2018){Kisaka}, {Ioka}, {Kashiyama}, and
  {Nakamura}}]{Kisaka2018}
{Kisaka} S, {Ioka} K, {Kashiyama} K, {Nakamura} T.
\newblock {Scattered Short Gamma-Ray Bursts as Electromagnetic Counterparts to
  Gravitational Waves and Implications of GW170817 and GRB 170817A}.
\newblock {\em \apj\/} {\bf 867} (2018) 39.
\newblock \doi{10.3847/1538-4357/aae30a}.

\bibitem[{{van Paradijs} et~al.(2000){van Paradijs}, {Kouveliotou}, and
  {Wijers}}]{vanParadijs2000}
{van Paradijs} J, {Kouveliotou} C, {Wijers} RAMJ.
\newblock {Gamma-Ray Burst Afterglows}.
\newblock {\em \araa\/} {\bf 38} (2000) 379--425.
\newblock \doi{10.1146/annurev.astro.38.1.379}.

\bibitem[{{Nousek} et~al.(2006){Nousek}, {Kouveliotou}, {Grupe}, {Page},
  {Granot}, {Ramirez-Ruiz} et~al.}]{Nousek2006}
{Nousek} JA, {Kouveliotou} C, {Grupe} D, {Page} KL, {Granot} J, {Ramirez-Ruiz}
  E, et~al.
\newblock {Evidence for a Canonical Gamma-Ray Burst Afterglow Light Curve in
  the Swift XRT Data}.
\newblock {\em \apj\/} {\bf 642} (2006) 389--400.
\newblock \doi{10.1086/500724}.

\bibitem[{{Ruan} et~al.(2018){Ruan}, {Nynka}, {Haggard}, {Kalogera}, and
  {Evans}}]{Ruan2018}
{Ruan} JJ, {Nynka} M, {Haggard} D, {Kalogera} V, {Evans} P.
\newblock {Brightening X-Ray Emission from GW170817/GRB 170817A: Further
  Evidence for an Outflow}.
\newblock {\em \apjl\/} {\bf 853} (2018) L4.
\newblock \doi{10.3847/2041-8213/aaa4f3}.

\bibitem[{{Troja} et~al.(2019{\natexlab{a}}){Troja}, {van Eerten}, {Ryan},
  {Ricci}, {Burgess}, {Wieringa} et~al.}]{Troja2019}
{Troja} E, {van Eerten} H, {Ryan} G, {Ricci} R, {Burgess} JM, {Wieringa} MH,
  et~al.
\newblock {A year in the life of GW 170817: the rise and fall of a structured
  jet from a binary neutron star merger}.
\newblock {\em \mnras\/} {\bf 489} (2019{\natexlab{a}}) 1919--1926.
\newblock \doi{10.1093/mnras/stz2248}.

\bibitem[{{Dobie} et~al.(2018){Dobie}, {Kaplan}, {Murphy}, {Lenc}, {Mooley},
  {Lynch} et~al.}]{Dobie2018}
{Dobie} D, {Kaplan} DL, {Murphy} T, {Lenc} E, {Mooley} KP, {Lynch} C, et~al.
\newblock {A Turnover in the Radio Light Curve of GW170817}.
\newblock {\em \apjl\/} {\bf 858} (2018) L15.
\newblock \doi{10.3847/2041-8213/aac105}.

\bibitem[{{Makhathini} et~al.(2020){Makhathini}, {Mooley}, {Brightman},
  {Hotokezaka}, {Nayana}, {Intema} et~al.}]{Makhathini2020}
{Makhathini} S, {Mooley} KP, {Brightman} M, {Hotokezaka} K, {Nayana} A,
  {Intema} HT, et~al.
\newblock {The Panchromatic Afterglow of GW170817: The full uniform dataset,
  modeling, comparison with previous results and implications}.
\newblock {\em arXiv e-prints\/}  (2020) arXiv:2006.02382.

\bibitem[{{Fong} et~al.(2017){Fong}, {Berger}, {Blanchard}, {Margutti},
  {Cowperthwaite}, {Chornock} et~al.}]{Fong2017}
{Fong} W, {Berger} E, {Blanchard} PK, {Margutti} R, {Cowperthwaite} PS,
  {Chornock} R, et~al.
\newblock {The Electromagnetic Counterpart of the Binary Neutron Star Merger
  LIGO/Virgo GW170817. VIII. A Comparison to Cosmological Short-duration
  Gamma-Ray Bursts}.
\newblock {\em \apjl\/} {\bf 848} (2017) L23.
\newblock \doi{10.3847/2041-8213/aa9018}.

\bibitem[{{Lazzati} et~al.(2018){Lazzati}, {Perna}, {Morsony}, {Lopez-Camara},
  {Cantiello}, {Ciolfi} et~al.}]{Lazzati2018}
{Lazzati} D, {Perna} R, {Morsony} BJ, {Lopez-Camara} D, {Cantiello} M, {Ciolfi}
  R, et~al.
\newblock {Late Time Afterglow Observations Reveal a Collimated Relativistic
  Jet in the Ejecta of the Binary Neutron Star Merger GW170817}.
\newblock {\em \prl\/} {\bf 120} (2018) 241103.
\newblock \doi{10.1103/PhysRevLett.120.241103}.

\bibitem[{{Salafia} et~al.(2018){Salafia}, {Ghisellini}, {Ghirlanda}, and
  {Colpi}}]{Salafia2018}
{Salafia} OS, {Ghisellini} G, {Ghirlanda} G, {Colpi} M.
\newblock {Interpreting GRB170817A as a giant flare from a jet-less double
  neutron star merger}.
\newblock {\em \aap\/} {\bf 619} (2018) A18.
\newblock \doi{10.1051/0004-6361/201732259}.

\bibitem[{{Granot} et~al.(2018{\natexlab{b}}){Granot}, {Gill}, {Guetta}, and
  {De Colle}}]{Granot2018}
{Granot} J, {Gill} R, {Guetta} D, {De Colle} F.
\newblock {Off-axis emission of short GRB jets from double neutron star mergers
  and GRB 170817A}.
\newblock {\em \mnras\/} {\bf 481} (2018{\natexlab{b}}) 1597--1608.
\newblock \doi{10.1093/mnras/sty2308}.

\bibitem[{{Lamb} and {Kobayashi}(2018)}]{Lamb2018a}
{Lamb} GP, {Kobayashi} S.
\newblock {GRB 170817A as a jet counterpart to gravitational wave triggerGW
  170817}.
\newblock {\em \mnras\/} {\bf 478} (2018) 733--740.
\newblock \doi{10.1093/mnras/sty1108}.

\bibitem[{{Finstad} et~al.(2018){Finstad}, {De}, {Brown}, {Berger}, and
  {Biwer}}]{Finstad2018}
{Finstad} D, {De} S, {Brown} DA, {Berger} E, {Biwer} CM.
\newblock {Measuring the Viewing Angle of GW170817 with Electromagnetic and
  Gravitational Waves}.
\newblock {\em \apjl\/} {\bf 860} (2018) L2.
\newblock \doi{10.3847/2041-8213/aac6c1}.

\bibitem[{{Xiao} et~al.(2017){Xiao}, {Liu}, {Dai}, and {Wu}}]{Xiao2017}
{Xiao} D, {Liu} LD, {Dai} ZG, {Wu} XF.
\newblock {Afterglows and Kilonovae Associated with Nearby Low-luminosity
  Short-duration Gamma-Ray Bursts: Application to GW170817/GRB 170817A}.
\newblock {\em \apjl\/} {\bf 850} (2017) L41.
\newblock \doi{10.3847/2041-8213/aa9b2b}.

\bibitem[{{Oganesyan} et~al.(2020){Oganesyan}, {Ascenzi}, {Branchesi},
  {Salafia}, {Dall'Osso}, and {Ghirlanda}}]{Oganesyan2020}
{Oganesyan} G, {Ascenzi} S, {Branchesi} M, {Salafia} OS, {Dall'Osso} S,
  {Ghirlanda} G.
\newblock {Structured Jets and X-Ray Plateaus in Gamma-Ray Burst Phenomena}.
\newblock {\em \apj\/} {\bf 893} (2020) 88.
\newblock \doi{10.3847/1538-4357/ab8221}.

\bibitem[{{Fraija} et~al.(2019{\natexlab{a}}){Fraija}, {Pedreira}, and
  {Veres}}]{Fraija2019}
{Fraija} N, {Pedreira} ACCdES, {Veres} P.
\newblock {Light Curves of a Shock-breakout Material and a Relativistic
  Off-axis Jet from a Binary Neutron Star System}.
\newblock {\em \apj\/} {\bf 871} (2019{\natexlab{a}}) 200.
\newblock \doi{10.3847/1538-4357/aaf80e}.

\bibitem[{{Fraija} et~al.(2019{\natexlab{b}}){Fraija}, {De Colle}, {Veres},
  {Dichiara}, {Barniol Duran}, {Galvan-Gamez} et~al.}]{Fraija2019b}
{Fraija} N, {De Colle} F, {Veres} P, {Dichiara} S, {Barniol Duran} R,
  {Galvan-Gamez} A, et~al.
\newblock {The Short GRB 170817A: Modeling the Off-axis Emission and
  Implications on the Ejecta Magnetization}.
\newblock {\em \apj\/} {\bf 871} (2019{\natexlab{b}}) 123.
\newblock \doi{10.3847/1538-4357/aaf564}.

\bibitem[{{De Colle} et~al.(2018){De Colle}, {Kumar}, and
  {Aguilera-Dena}}]{Decolle2018}
{De Colle} F, {Kumar} P, {Aguilera-Dena} DR.
\newblock {Radio Emission from the Cocoon of a GRB Jet: Implications for
  Relativistic Supernovae and Off-axis GRB Emission}.
\newblock {\em \apj\/} {\bf 863} (2018) 32.
\newblock \doi{10.3847/1538-4357/aad04d}.

\bibitem[{{Nakar} and {Piran}(2018)}]{Nakar2018}
{Nakar} E, {Piran} T.
\newblock {Implications of the radio and X-ray emission that followed
  GW170817}.
\newblock {\em \mnras\/} {\bf 478} (2018) 407--415.
\newblock \doi{10.1093/mnras/sty952}.

\bibitem[{{Li} et~al.(2018){Li}, {Li}, {Huang}, {Geng}, {Yu}, and
  {Song}}]{Li2018}
{Li} B, {Li} LB, {Huang} YF, {Geng} JJ, {Yu} YB, {Song} LM.
\newblock {Continued Brightening of the Afterglow of GW170817/GRB 170817A as
  Being Due to a Delayed Energy Injection}.
\newblock {\em \apjl\/} {\bf 859} (2018) L3.
\newblock \doi{10.3847/2041-8213/aac2c5}.

\bibitem[{{Alexander} et~al.(2018){Alexander}, {Margutti}, {Blanchard}, {Fong},
  {Berger}, {Hajela} et~al.}]{Alexander2018}
{Alexander} KD, {Margutti} R, {Blanchard} PK, {Fong} W, {Berger} E, {Hajela} A,
  et~al.
\newblock {A Decline in the X-Ray through Radio Emission from GW170817
  Continues to Support an Off-axis Structured Jet}.
\newblock {\em \apjl\/} {\bf 863} (2018) L18.
\newblock \doi{10.3847/2041-8213/aad637}.

\bibitem[{{Mooley} et~al.(2018{\natexlab{c}}){Mooley}, {Frail}, {Dobie},
  {Lenc}, {Corsi}, {De} et~al.}]{Mooley2018c}
{Mooley} KP, {Frail} DA, {Dobie} D, {Lenc} E, {Corsi} A, {De} K, et~al.
\newblock {A Strong Jet Signature in the Late-time Light Curve of GW170817}.
\newblock {\em \apjl\/} {\bf 868} (2018{\natexlab{c}}) L11.
\newblock \doi{10.3847/2041-8213/aaeda7}.

\bibitem[{{Nynka} et~al.(2018){Nynka}, {Ruan}, {Haggard}, and
  {Evans}}]{Nynka2018}
{Nynka} M, {Ruan} JJ, {Haggard} D, {Evans} PA.
\newblock {Fading of the X-Ray Afterglow of Neutron Star Merger GW170817/GRB
  170817A at 260 Days}.
\newblock {\em \apjl\/} {\bf 862} (2018) L19.
\newblock \doi{10.3847/2041-8213/aad32d}.

\bibitem[{{Hajela} et~al.(2019){Hajela}, {Margutti}, {Alexander},
  {Kathirgamaraju}, {Baldeschi}, {Guidorzi} et~al.}]{Hajela2019}
{Hajela} A, {Margutti} R, {Alexander} KD, {Kathirgamaraju} A, {Baldeschi} A,
  {Guidorzi} C, et~al.
\newblock {Two Years of Nonthermal Emission from the Binary Neutron Star Merger
  GW170817: Rapid Fading of the Jet Afterglow and First Constraints on the
  Kilonova Fastest Ejecta}.
\newblock {\em \apjl\/} {\bf 886} (2019) L17.
\newblock \doi{10.3847/2041-8213/ab5226}.

\bibitem[{{Lamb} et~al.(2018){Lamb}, {Mandel}, and {Resmi}}]{Lamb2018c}
{Lamb} GP, {Mandel} I, {Resmi} L.
\newblock {Late-time evolution of afterglows from off-axis neutron star
  mergers}.
\newblock {\em \mnras\/} {\bf 481} (2018) 2581--2589.
\newblock \doi{10.1093/mnras/sty2196}.

\bibitem[{{Fong} et~al.(2019){Fong}, {Blanchard}, {Alexander}, {Strader},
  {Margutti}, {Hajela} et~al.}]{Fong2019}
{Fong} W, {Blanchard} PK, {Alexander} KD, {Strader} J, {Margutti} R, {Hajela}
  A, et~al.
\newblock {The Optical Afterglow of GW170817: An Off-axis Structured Jet and
  Deep Constraints on a Globular Cluster Origin}.
\newblock {\em \apjl\/} {\bf 883} (2019) L1.
\newblock \doi{10.3847/2041-8213/ab3d9e}.

\bibitem[{{Jin} et~al.(2018){Jin}, {Li}, {Wang}, {Wang}, {He}, {Yuan}
  et~al.}]{Jin2018}
{Jin} ZP, {Li} X, {Wang} H, {Wang} YZ, {He} HN, {Yuan} Q, et~al.
\newblock {Short GRBs: Opening Angles, Local Neutron Star Merger Rate, and
  Off-axis Events for GRB/GW Association}.
\newblock {\em \apj\/} {\bf 857} (2018) 128.
\newblock \doi{10.3847/1538-4357/aab76d}.

\bibitem[{{Gill} and {Granot}(2018)}]{Gill2018}
{Gill} R, {Granot} J.
\newblock {Afterglow imaging and polarization of misaligned structured GRB jets
  and cocoons: breaking the degeneracy in GRB 170817A}.
\newblock {\em \mnras\/} {\bf 478} (2018) 4128--4141.
\newblock \doi{10.1093/mnras/sty1214}.

\bibitem[{{Corsi} et~al.(2018){Corsi}, {Hallinan}, {Lazzati}, {Mooley},
  {Murphy}, {Frail} et~al.}]{Corsi2018}
{Corsi} A, {Hallinan} GW, {Lazzati} D, {Mooley} KP, {Murphy} EJ, {Frail} DA,
  et~al.
\newblock {An Upper Limit on the Linear Polarization Fraction of the GW170817
  Radio Continuum}.
\newblock {\em \apjl\/} {\bf 861} (2018) L10.
\newblock \doi{10.3847/2041-8213/aacdfd}.

\bibitem[{{Yamazaki} et~al.(2018){Yamazaki}, {Ioka}, and
  {Nakamura}}]{Yamazaki2018}
{Yamazaki} R, {Ioka} K, {Nakamura} T.
\newblock {Prompt emission from the counter jet of a short gamma-ray burst}.
\newblock {\em Progress of Theoretical and Experimental Physics\/} {\bf 2018}
  (2018) 033E01.
\newblock \doi{10.1093/ptep/pty012}.

\bibitem[{{M{\'e}sz{\'a}ros} and {Rees}(1997)}]{Meszaros1997}
{M{\'e}sz{\'a}ros} P, {Rees} MJ.
\newblock {Optical and Long-Wavelength Afterglow from Gamma-Ray Bursts}.
\newblock {\em \apj\/} {\bf 476} (1997) 232--237.
\newblock \doi{10.1086/303625}.

\bibitem[{{Sari} et~al.(1998){Sari}, {Piran}, and {Narayan}}]{Sari1998}
{Sari} R, {Piran} T, {Narayan} R.
\newblock {Spectra and Light Curves of Gamma-Ray Burst Afterglows}.
\newblock {\em \apjl\/} {\bf 497} (1998) L17--L20.
\newblock \doi{10.1086/311269}.

\bibitem[{{Granot} et~al.(2002){Granot}, {Panaitescu}, {Kumar}, and
  {Woosley}}]{Granot2002}
{Granot} J, {Panaitescu} A, {Kumar} P, {Woosley} SE.
\newblock {Off-Axis Afterglow Emission from Jetted Gamma-Ray Bursts}.
\newblock {\em \apjl\/} {\bf 570} (2002) L61--L64.
\newblock \doi{10.1086/340991}.

\bibitem[{{Troja} et~al.(2018{\natexlab{a}}){Troja}, {Piro}, {Ryan}, {van
  Eerten}, {Ricci}, {Wieringa} et~al.}]{Troja2018}
{Troja} E, {Piro} L, {Ryan} G, {van Eerten} H, {Ricci} R, {Wieringa} MH, et~al.
\newblock {The outflow structure of GW170817 from late-time broad-band
  observations}.
\newblock {\em \mnras\/} {\bf 478} (2018{\natexlab{a}}) L18--L23.
\newblock \doi{10.1093/mnrasl/sly061}.

\bibitem[{{Beniamini} and {Nakar}(2019)}]{Beniamini2019a}
{Beniamini} P, {Nakar} E.
\newblock {Observational constraints on the structure of gamma-ray burst jets}.
\newblock {\em \mnras\/} {\bf 482} (2019) 5430--5440.
\newblock \doi{10.1093/mnras/sty3110}.

\bibitem[{{Salafia} et~al.(2019){Salafia}, {Ghirlanda}, {Ascenzi}, and
  {Ghisellini}}]{Salafia2019}
{Salafia} OS, {Ghirlanda} G, {Ascenzi} S, {Ghisellini} G.
\newblock {On-axis view of GRB 170817A}.
\newblock {\em \aap\/} {\bf 628} (2019) A18.
\newblock \doi{10.1051/0004-6361/201935831}.

\bibitem[{{Mandel}(2018)}]{Mandel2018}
{Mandel} I.
\newblock {The Orbit of GW170817 Was Inclined by Less Than 28$^o$ to the Line
  of Sight}.
\newblock {\em \apjl\/} {\bf 853} (2018) L12.
\newblock \doi{10.3847/2041-8213/aaa6c1}.

\bibitem[{{Zou} et~al.(2018){Zou}, {Wang}, {Moharana}, {Liao}, {Chen}, {Wu}
  et~al.}]{Zou2018}
{Zou} YC, {Wang} FF, {Moharana} R, {Liao} B, {Chen} W, {Wu} Q, et~al.
\newblock {Determining the Lorentz Factor and Viewing Angle of GRB 170817A}.
\newblock {\em \apjl\/} {\bf 852} (2018) L1.
\newblock \doi{10.3847/2041-8213/aaa123}.

\bibitem[{{Metzger} et~al.(2018){Metzger}, {Thompson}, and
  {Quataert}}]{Metzger2018}
{Metzger} BD, {Thompson} TA, {Quataert} E.
\newblock {A Magnetar Origin for the Kilonova Ejecta in GW170817}.
\newblock {\em \apj\/} {\bf 856} (2018) 101.
\newblock \doi{10.3847/1538-4357/aab095}.

\bibitem[{{Piro} et~al.(2019){Piro}, {Troja}, {Zhang}, {Ryan}, {van Eerten},
  {Ricci} et~al.}]{Piro2019}
{Piro} L, {Troja} E, {Zhang} B, {Ryan} G, {van Eerten} H, {Ricci} R, et~al.
\newblock {A long-lived neutron star merger remnant in GW170817: constraints
  and clues from X-ray observations}.
\newblock {\em \mnras\/} {\bf 483} (2019) 1912--1921.
\newblock \doi{10.1093/mnras/sty3047}.

\bibitem[{{Pooley} et~al.(2018){Pooley}, {Kumar}, {Wheeler}, and
  {Grossan}}]{Pooley2018}
{Pooley} D, {Kumar} P, {Wheeler} JC, {Grossan} B.
\newblock {GW170817 Most Likely Made a Black Hole}.
\newblock {\em \apjl\/} {\bf 859} (2018) L23.
\newblock \doi{10.3847/2041-8213/aac3d6}.

\bibitem[{{Abedi} and {Afshordi}(2019)}]{Abedi2019}
{Abedi} J, {Afshordi} N.
\newblock {Echoes from the abyss: a highly spinning black hole remnant for the
  binary neutron star merger GW170817}.
\newblock {\em \jcap\/} {\bf 2019} (2019) 010.
\newblock \doi{10.1088/1475-7516/2019/11/010}.

\bibitem[{{Kyutoku} et~al.(2020){Kyutoku}, {Fujibayashi}, {Hayashi},
  {Kawaguchi}, {Kiuchi}, {Shibata} et~al.}]{Kyutoku2020}
{Kyutoku} K, {Fujibayashi} S, {Hayashi} K, {Kawaguchi} K, {Kiuchi} K, {Shibata}
  M, et~al.
\newblock {On the Possibility of GW190425 Being a Black Hole-Neutron Star
  Binary Merger}.
\newblock {\em \apjl\/} {\bf 890} (2020) L4.
\newblock \doi{10.3847/2041-8213/ab6e70}.

\bibitem[{{Beniamini} et~al.(2019){Beniamini}, {Petropoulou}, {Barniol Duran},
  and {Giannios}}]{Beniamini2019}
{Beniamini} P, {Petropoulou} M, {Barniol Duran} R, {Giannios} D.
\newblock {A lesson from GW170817: most neutron star mergers result in tightly
  collimated successful GRB jets}.
\newblock {\em \mnras\/} {\bf 483} (2019) 840--851.
\newblock \doi{10.1093/mnras/sty3093}.

\bibitem[{{Troja} et~al.(2018{\natexlab{b}}){Troja}, {Ryan}, {Piro}, {van
  Eerten}, {Cenko}, {Yoon} et~al.}]{Troja2018a}
{Troja} E, {Ryan} G, {Piro} L, {van Eerten} H, {Cenko} SB, {Yoon} Y, et~al.
\newblock {A luminous blue kilonova and an off-axis jet from a compact binary
  merger at z = 0.1341}.
\newblock {\em Nature Communications\/} {\bf 9} (2018{\natexlab{b}}) 4089.
\newblock \doi{10.1038/s41467-018-06558-7}.

\bibitem[{{Troja} et~al.(2019{\natexlab{b}}){Troja}, {Castro-Tirado}, {Becerra
  Gonz{\'a}lez}, {Hu}, {Ryan}, {Cenko} et~al.}]{Troja2019b}
{Troja} E, {Castro-Tirado} AJ, {Becerra Gonz{\'a}lez} J, {Hu} Y, {Ryan} GS,
  {Cenko} SB, et~al.
\newblock {The afterglow and kilonova of the short GRB 160821B}.
\newblock {\em \mnras\/} {\bf 489} (2019{\natexlab{b}}) 2104--2116.
\newblock \doi{10.1093/mnras/stz2255}.

\bibitem[{{Lamb} et~al.(2019{\natexlab{b}}){Lamb}, {Tanvir}, {Levan}, {de
  Ugarte Postigo}, {Kawaguchi}, {Corsi} et~al.}]{Lamb2019b}
{Lamb} GP, {Tanvir} NR, {Levan} AJ, {de Ugarte Postigo} A, {Kawaguchi} K,
  {Corsi} A, et~al.
\newblock {Short GRB 160821B: A Reverse Shock, a Refreshed Shock, and a
  Well-sampled Kilonova}.
\newblock {\em \apj\/} {\bf 883} (2019{\natexlab{b}}) 48.
\newblock \doi{10.3847/1538-4357/ab38bb}.

\bibitem[{{Burns} et~al.(2018){Burns}, {Veres}, {Connaughton}, {Racusin},
  {Briggs}, {Christensen} et~al.}]{Burns2018}
{Burns} E, {Veres} P, {Connaughton} V, {Racusin} J, {Briggs} MS, {Christensen}
  N, et~al.
\newblock {Fermi GBM Observations of GRB 150101B: A Second Nearby Event with a
  Short Hard Spike and a Soft Tail}.
\newblock {\em \apjl\/} {\bf 863} (2018) L34.
\newblock \doi{10.3847/2041-8213/aad813}.

\bibitem[{{von Kienlin} et~al.(2019){von Kienlin}, {Veres}, {Roberts},
  {Hamburg}, {Bissaldi}, {Briggs} et~al.}]{vonKienlin2019}
{von Kienlin} A, {Veres} P, {Roberts} OJ, {Hamburg} R, {Bissaldi} E, {Briggs}
  MS, et~al.
\newblock {Fermi-GBM GRBs with Characteristics Similar to GRB 170817A}.
\newblock {\em \apj\/} {\bf 876} (2019) 89.
\newblock \doi{10.3847/1538-4357/ab10d8}.

\end{thebibliography}

%%% Make sure to upload the bib file along with the tex file and PDF
%%% Please see the test.bib file for some examples of references

\section*{Figure captions}

%%% Please be aware that for original research articles we only permit a combined number of 15 figures and tables, one figure with multiple subfigures will count as only one figure.
%%% Use this if adding the figures directly in the mansucript, if so, please remember to also upload the files when submitting your article
%%% There is no need for adding the file termination, as long as you indicate where the file is saved. In the examples below the files (logo1.eps and logos.eps) are in the Frontiers LaTeX folder
%%% If using *.tif files convert them to .jpg or .png
%%%  NB logo1.eps is required in the path in order to correctly compile front page header %%%

%%% If you are submitting a figure with subfigures please combine these into one image file with part labels integrated.
%%% If you don't add the figures in the LaTeX files, please upload them when submitting the article.
%%% Frontiers will add the figures at the end of the provisional pdf automatically
%%% The use of LaTeX coding to draw Diagrams/Figures/Structures should be avoided. They should be external callouts including graphics.

\end{document}